\documentclass[aip,pof,amsmath,amssymb]{revtex4-1}
\usepackage{hyperref}

\usepackage{graphicx,subfigure,color,booktabs}

\newcommand{\bi}{ \boldsymbol}

\DeclareMathOperator{\sech}{sech}

\begin{document}

\title{Sloshing instability and electrolyte layer rupture in liquid metal batteries}

\author{Norbert Weber}
  \altaffiliation{Helmholtz-Zentrum Dresden - Rossendorf, Bautzner
    Landstr. 400, 01328 Dresden, Germany}

\author{Pascal Beckstein}
  \altaffiliation{Helmholtz-Zentrum Dresden - Rossendorf, Bautzner
    Landstr. 400, 01328 Dresden, Germany}

\author{Wietze Herreman}
  \altaffiliation{LIMSI, CNRS, Univ. Paris-Sud, Universit\'e Paris-Saclay, 
B\^at 508, rue John von Neumann, Campus Universitaire, F-91405 Orsay, France}

\author{Gerrit Maik Horstmann}
  \altaffiliation{Helmholtz-Zentrum Dresden - Rossendorf, Bautzner
    Landstr. 400, 01328 Dresden, Germany}

\author{Caroline Nore}
  \altaffiliation{LIMSI, CNRS, Univ. Paris-Sud, Universit\'e Paris-Saclay, 
B\^at 508, rue John von Neumann, Campus Universitaire, F-91405 Orsay, France}

\author{Frank Stefani}
  \altaffiliation{Helmholtz-Zentrum Dresden - Rossendorf, Bautzner
    Landstr. 400, 01328 Dresden, Germany}

\author{Tom Weier}
  \altaffiliation{Helmholtz-Zentrum Dresden - Rossendorf, Bautzner
    Landstr. 400, 01328 Dresden, Germany}

\date{\today}

\begin{abstract}
Liquid metal batteries (LMBs) are discussed today as a cheap grid scale
energy storage, as required for the deployment of fluctuating
renewable energies. Built as a stable density stratification of two
liquid metals separated by a thin molten salt layer, LMBs are susceptible to
short-circuit by fluid flows. Using direct numerical simulation, we
study a sloshing long wave interface instability in cylindrical cells, 
which is already known from aluminium reduction cells. After characterising the instability
mechanism, we investigate the influence of cell current, layer thickness,
density, viscosity, conductivity and magnetic background
field. Finally we study the shape of the interface and give a dimensionless parameter for the onset of
sloshing as well as for the short-circuit.
\end{abstract}

% http://www.aip.org/publishing/pacs/pacs-reg40#47
% 47.11.Df Finite volume methods
% 47.20.-k      Flow instabilities
% 52.30.Cv Magnetohydrodynamics
% 47.35.Tv 	Magnetohydrodynamic waves
\pacs{47.11.Df 47.20.-k 52.30.Cv 47.35.Tv}

\keywords{liquid metal battery, magnetohydrodynamics, flow
  instability, metal pad roll instability, sloshing}

\maketitle

%\linenumbers

\section{Introduction} \label{introduction}

Liquid metal batteries (LMBs) are among the systems currently discussed for
electrochemical energy storage on the grid level. With growing share
of renewable and volatile sources, as wind and solar, demand for economic
large-scale storage rises. Future energy systems based primarily
or even exclusively on renewables can hardly be imagined without
adequate storage capacity if electricity demand has to be met
independently of the current weather conditions
\cite{Wittingham2012} and if grid stability shall be maintained
\cite{Nardelli2014}.

\begin{figure}[t!]
\centering\hspace{15mm}
 \subfigure[]{\includegraphics[width=0.25\textwidth]{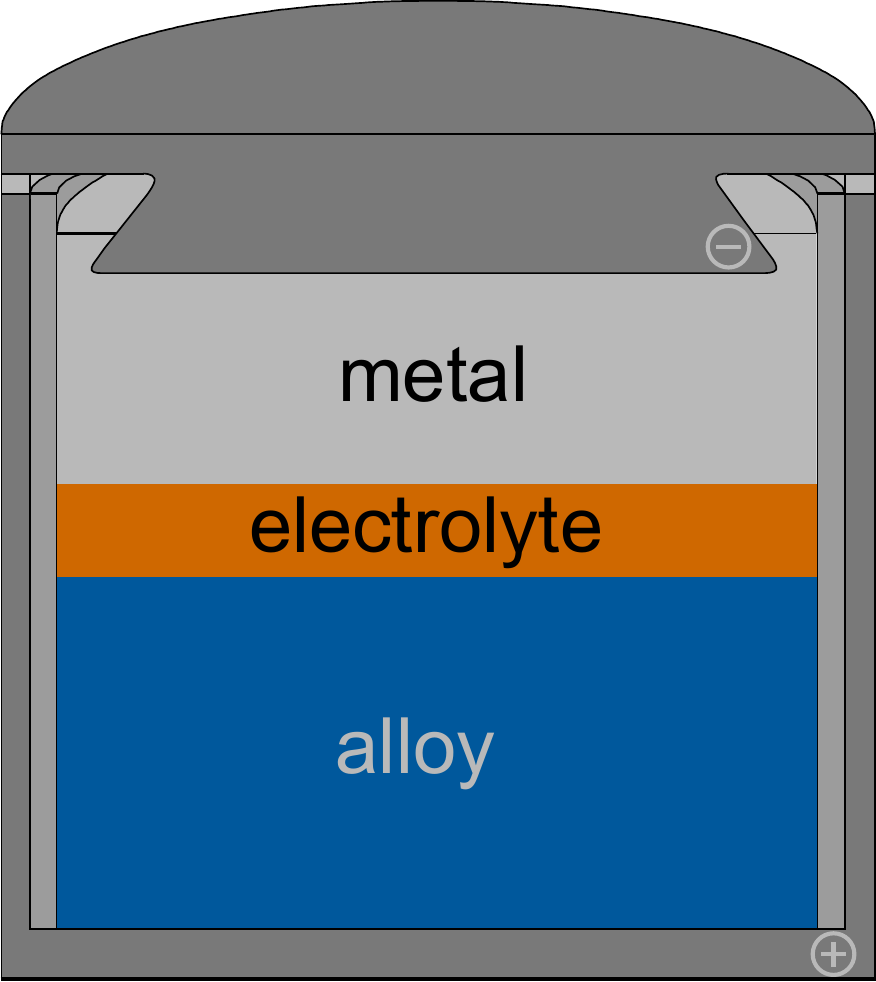}}\hfill
 \subfigure[]{\includegraphics[width=0.25\textwidth]{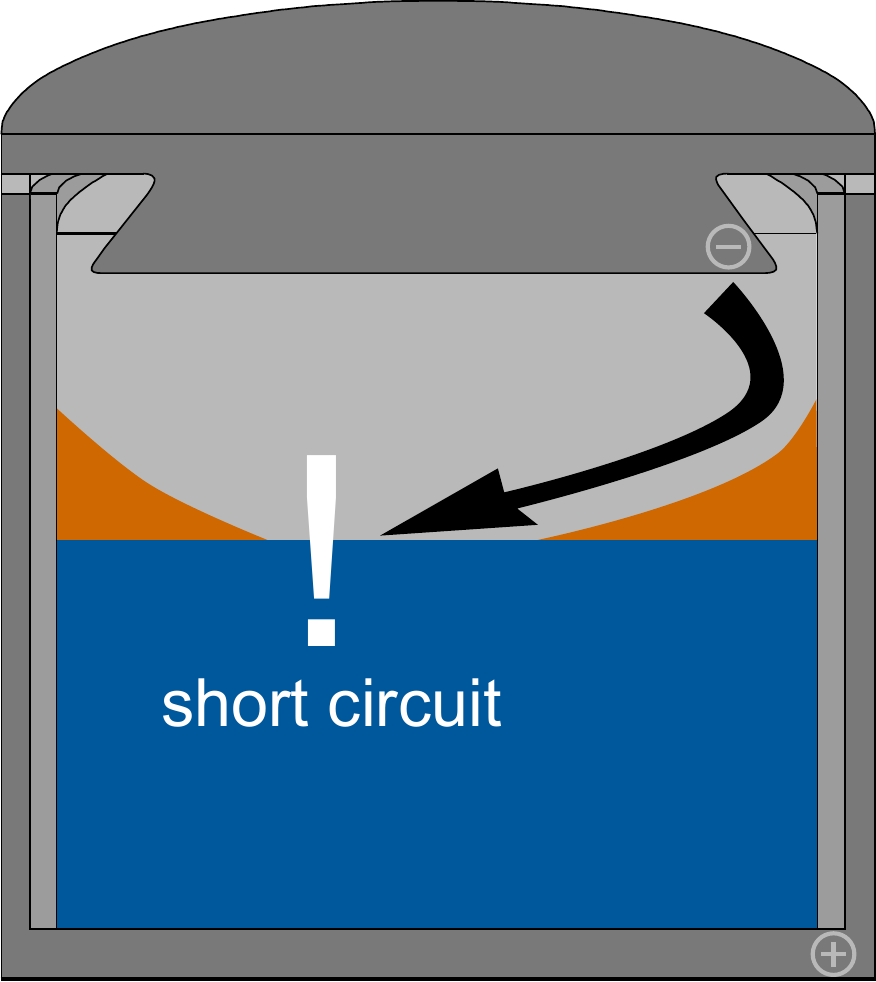}}\hspace{15mm}
 \caption{Sketch of a liquid metal battery (a) and short-circuit due
to a deformation of the electrolyte layer (b).}
\label{f:sketchLMB}
\end{figure}

LMBs consist of a stable density stratification of three 
liquids: a low density alkaline or earth-alkaline liquid metal on the
top, a heavy metal on the bottom and a medium density molten salt
mixture sandwiched in between (``differential density cell''
\cite{Agruss1967}; see figure \ref{f:sketchLMB}a). The operation
temperature lies slightly above the highest melting point of the
active materials (typically between 275 and 700\,$^\circ$C).
Originally, LMBs were investigated as part of
thermally regenerative energy conversion systems
\cite{Weaver1962,Agruss1963}, but focus of research later shifted to
their application as electricity storage devices
\cite{Steunenberg2000}. Progress in the field during the 1960s and
early 1970s has been reviewed, e.g., by Crouthamel and Recht
\cite{Crouthamel1967}, Cairns et al. \cite{Cairns1967}, Cairns and
Shimotake \cite{Cairns1969b,Swinkels1971}, and recently by Kim
et al. \cite{Kim2013b}. According to the latter authors, research came
to a halt in the 1970s because the low specific energy of LMBs rendered
them unattractive for portable applications and ``much of the
aforementioned research fell into obscurity for the next few
decades''.

Interest in LMBs has been recently renewed, sparked by the work of
Donald Sadoway and his group at MIT\cite{Kim2013b}. The focus is now on cost-driven
development \cite{Spatocco2015} and grid-scale electrochemical
storage \cite{Kim2013b}. Different active material combinations and
electrolytes are currently under investigation, ranging from Mg$||$Sb
\cite{Bradwell2012}, Ca$||$Bi \cite{Kim2013a}, Ca$||$Sb
\cite{Ouchi2014}, Ca-Mg$||$Bi \cite{Ouchi2016}, Li$||$Sb-Pb
\cite{Wang2014} to Na$||$Pb-Bi \cite{Spatocco2015a}.

Due to their completely liquid interior, LMBs have attracted the
attention of fluid dynamists as well. A number of recent publications
are devoted to the problem of the Tayler instability and its
circumvention \cite{Stefani2011,Weber2014,Herreman2015}, to temperature-driven
convection \cite{Shen2015,Perez2015,Beltran2016}  and electro-vortex flows
\cite{Kelley2014,Weber2014b,Stefani2015} in LMBs, and to a simplified
model of sloshing in a three layer system \cite{Zikanov2015}. These
investigations are motivated on the one hand by the need to prevent a
direct contact between anode and cathode melt that could occur if
violent motion would develop in the electrode(s) (see
figure \ref{f:sketchLMB}b). On the other hand, mass transport in the
lower metal is often limiting the cell performance \cite{Heredy1967,Agruss1967,Bradwell2012,Kim2013b,Kim2013a,Ouchi2014,Ouchi2016}
 calling for enhanced mixing. This could at the same time
prevent the sometimes observed accumulation of intermetallic compounds
\cite{Shimotake1967,Kim2012,Kim2013a}
at the electrode-electrolyte interface.
 
LMBs are thought to be easily scalable on the cell level due to their simple
construction and the self-assembling of the liquid layers. Large cells
(in the order of cubic meters \cite{Bradwell2011,Sadoway2012}) are supposed to
operate at very high power values \cite{Huggins2016}. Current
densities of up to 13\,A/cm$^2$ were measured for Li$||$Te cells
\cite{Cairns1969b}, and less exotic couples as Na$||$Bi\cite{Shimotake1967} still reach
1\,A/cm$^2$. Together, high current densities
and large electrode areas result in strong total currents that are
able to generate significant electromagnetic forces. Such forces may
give rise to the aforementioned Tayler instability, but may also
generate a long wave interfacial instability known from aluminium
reduction cells as ``sloshing'' or ``metal pad roll instability''. The
manifestation of this instability in LMBs is the topic of the paper at
hand.

\begin{figure}[t!]
\centering
 \subfigure[]{\includegraphics[height=0.3\textwidth]{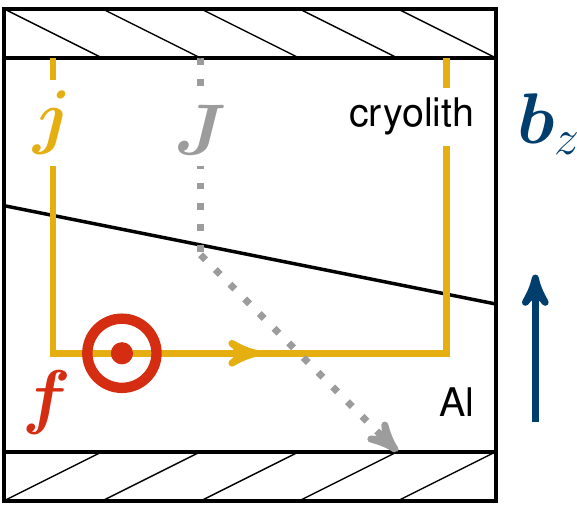}}\hfill
 \subfigure[]{\includegraphics[height=0.3\textwidth]{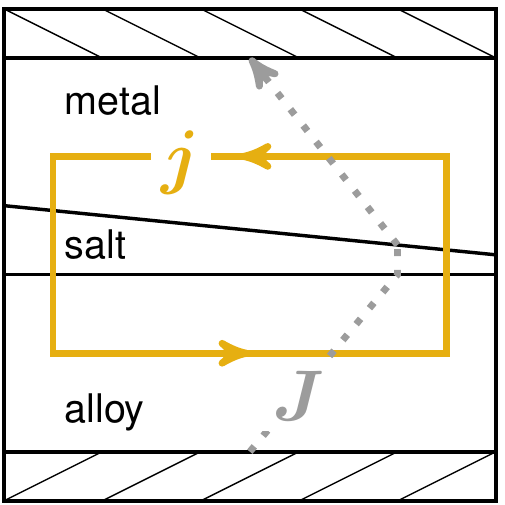}}\hfill
\caption{Cross section of an aluminium reduction cell (a) and a liquid metal
  battery (b) with tilted interface \cite{Sele1977,Davidson1998}. $\bi
  J$ denotes the total cell current, 
$\bi j$ the perturbed or compensation current, ${\bi B_{0,z}}$ a vertical
magnetic field and $\bi f$ the resulting Lorentz force.}
\label{f:mpr:alu}
\end{figure}

As just mentioned, the metal pad roll instability is well known from
the Hall-H\'eroult process of aluminium production. This two-phase
system consists of a stable density stratification of a molten salt
mixture (cryolite) floating atop liquid aluminium (figure
\ref{f:mpr:alu}a). A vertical current is applied by graphite current
collectors in order to reduce Al$_2$O$_3$ (solved in the cryolite) to
Al. Although this system works with only two liquid phases, it is
quite similar to liquid metal batteries: it operates at about
1\,000\,$^\circ$C, the density difference is only 200\,kg/m$^3$, the
salt resistivity is four orders of magnitude higher than the one of
the metal, the current density\cite{Evans2007} may reach
1\,A/cm$^2$. The main difference is the geometry: aluminium
reduction cells are typically rectangular (4\,x\,10\,m$^2$) and shallow 
\cite{Gerbeau2006,Davidson2000,Molokov2011,Pedchenko2009}.

\begin{figure}[t!]
\centering
\includegraphics[width=0.5\textwidth]{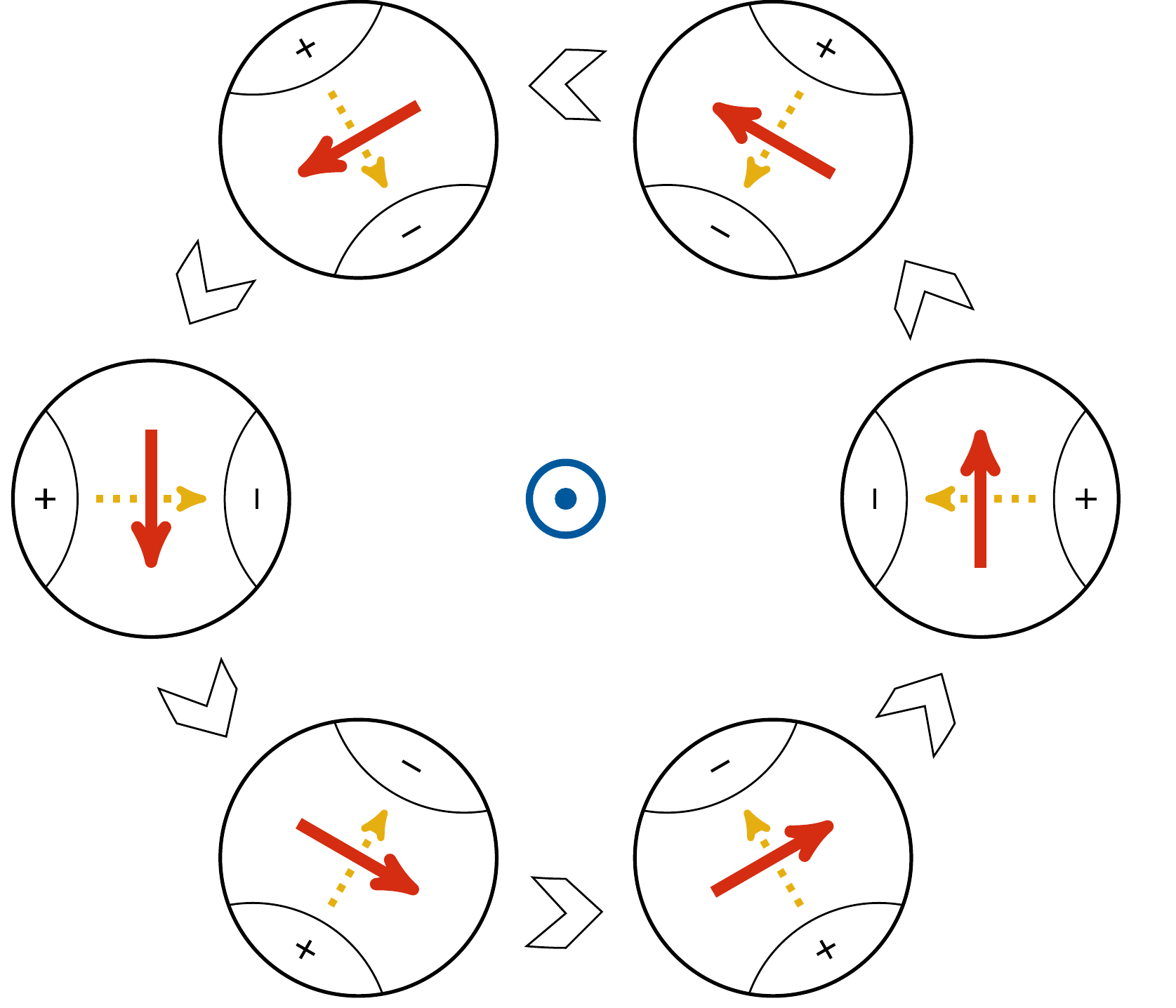}
\caption{Top view of the rotation of a tilted interface in a hypotetical aluminium reduction
  cell with circular shape\cite{Munger2008}. The compensation current (yellow) flows from a crest (+) to a
  trough (-), the Lorentz force (red) is orthogonal to current (yellow) and
  magnetic field (blue). For the orientation of global current and magnetic field,
  see figure \ref{f:mpr:alu}.}
\label{f:mpr:topview}
\end{figure}

A possible mechanism explaining the origin of metal pad rolling in
aluminium reduction cells was first explained by Sele
\cite{Sele1977}. We consider a Hall-H\'eroult cell with
a slightly inclined interface between aluminium and cryolite, see
figure \ref{f:mpr:alu}a. The electrolysis current $\bi J$ will 
take the easy way -- this is where the salt layer is thin. A
deformation of the interface will thus lead to a perturbed, or
compensation current $\bi j$ with a horizontal component.
The main idea of Sele's
model is an interaction of this \emph{horizontal} current
with a \emph{vertical} magnetic field $\bi B_{0,z}$. The latter one originates e.g. from
the supply lines; its typical value is between 1 and 10 mT\cite{Molokov2011}. The cross
product of horizontal current and vertical field, the Lorentz force
$\bi f$ is pointing towards the observer. 
Considering only the profile of
the cell (figure \ref{f:mpr:alu}a) it is not so obvious why the
Lorentz force will lead to a rotating wave. We illustrate therefore in
figure \ref{f:mpr:topview} the tilted interface for six different time
steps. The perturbed current flows always from a crest (+) of aluminium to a trough
(-); the Lorentz force is orthogonal. If we assume the Lorentz force
to displace only the crest, we can understand how the rotation
develops.

The instability mechanism described above may also be applied to a three
layer system of a liquid metal battery \cite{Weber2016b}. In order to
understand certain 
differences, it is important to know the electric conductivity of the
phases: for the salt it is low ($\sim$10$^2$\,S/m), high for the current
collectors ($\sim$10$^5$\,S/m) and even higher for the liquid metals
($\sim$10$^6$\,S/m). The strong resistance of the molten salt leads 
to a purely vertical current in the electrolyte layer (see figure 
\ref{f:mpr:alu}a and b). In an aluminium reduction cell, the perturbed
current has therefore to close in the current collector. In an LMB,
it closes already in the upper (liquid) electrode, leading to an
additional Lorentz force compared to aluminium reduction cells.
This mechanism will be explained in more detail in chapter
\ref{ch:instability_mechanism}. 

Denoting by $\bi J_0$ the current density of an unperturbed cell, the
sense of the flow can easily be determined by a simple rule. If $\bi
J_0\cdot \bi B_{0,z} > 0$,
the liquid metal layer at the bottom will rotate clockwise; if $\bi J_0\cdot \bi B_{0,z} < 0$
 the flow in the lower metal will be anti-clockwise. This holds for aluminium reduction
cells as well as liquid metal batteries. The upper metal layer in LMBs 
will flow in the opposite direction as the bottom metal. 
Only the upper interface
will deform notably -- therefore the wave rotates in the same 
direction as the upper metal.
\begin{figure}[t!]
\centering\hspace{5mm}
 \subfigure[]{\includegraphics[height=0.35\textwidth]{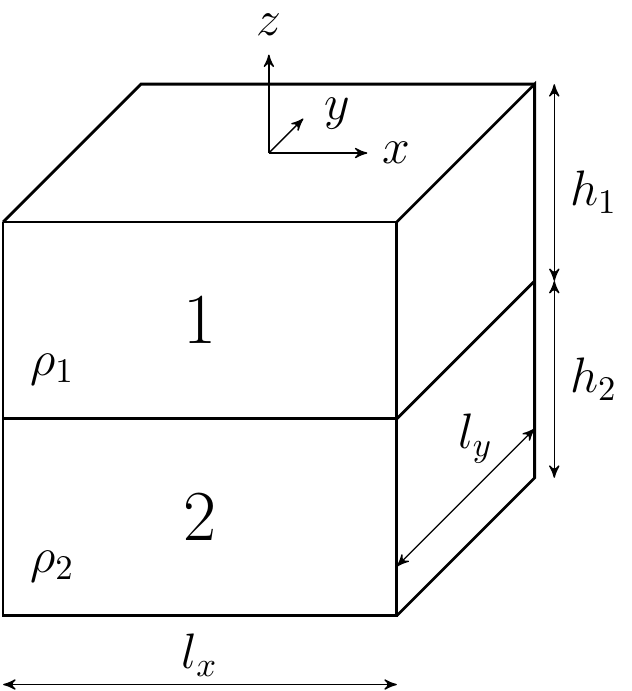}}\hfill
 \subfigure[]{\includegraphics[height=0.35\textwidth]{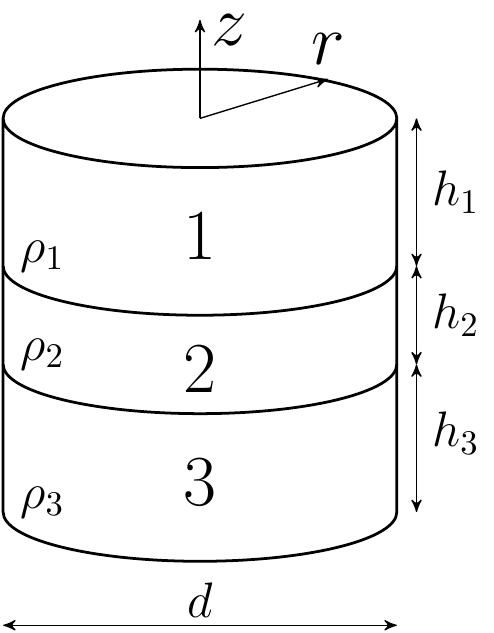}}\hspace{5mm}
\caption{Sketch of a an aluminium reduction cell (a) and liquid metal
  battery (b).}
\label{f:dimensions}
\end{figure}

The onset criterion of metal pad rolling in (rectangular) aluminium reduction 
cells was first described by Sele\cite{Sele1977,Molokov2011} as
\begin{equation}\label{eqn:beta:sele}
\beta = \frac{J_0 B_{0,z}}{g\Delta\rho}\cdot\frac{l_x}{h_1}\cdot\frac{l_y}{h_2} > \beta_\text{cr},
%  \beta = \frac{I B_{0,z}}{g\Delta\rho h_1 h_2} > \beta_\text{cr},
\end{equation}
i.e. for stable operation, the dimensionless number $\beta$ must
not exceed a certain critical value\cite{Molokov2011} in the order of $\beta_\text{cr}=1\dots340$. 
We will denote by $I$, $J_0$ and $B_{0,z}$ the absolute
values of cell current, current density and magnetic background field; 
$g$, $\Delta\rho=\rho_2-\rho_1$, $h_1$, $h_2$, $l_x$ and $l_y$ refer 
to the gravity, density difference, the height of the electrolyte 
and aluminium layer and the lateral dimensions of the aluminium
reduction cell, respectively 
(see also figure \ref{f:dimensions}a). The first factor in the definition
of $\beta$ is the relation of Lorentz force to gravity force, the 
two others are aspect ratios  and account for the layer thickness.
%For arbitrary geometries, $\beta$ is often expressed as $\beta = \frac{I B_{0,z}}{g\Delta
%\rho h_1 h_2}$.

%Whenever $\beta$ is high,
%the instability is more likely to occur. This means for liquid metal
%batteries: 
%\begin{itemize}
%\item flat cells are unstable
%\item big cells with large currents are unstable
%\item cells of similar density of upper metal and electrolyte are unstable
%\end{itemize}

Using a wave equation approach \cite{Urata1985}, a formula for $\beta_\text{cr}$
was later found as 
\begin{equation}\label{eqn:beta:bojarevics}
  \beta_\text{cr} = \pi^2 \left| m^2\frac{l_y}{l_x} - n^2\frac{l_x}{l_y}\right|
\end{equation}
by Bojarevics and Romerio
\cite{Bojarevics1994}. Here, $m$ and $n$ denote the wave number in $x$
and $y$ of a rectangular cell. Basically, this formula adds the
influence of the aspect ratio to the Sele criterion. As a main
consequence, square or cylindrical cells will always be unstable,
because $l_x=l_y$. 

Besides the vertical (and constant) magnetic background field $B_{0,z}$,
also spatial gradients of magnetic fields were suspected to lead to
instability \cite{Sneyd1992,Sneyd1994,Segatz1994}. The derived stability
condition for such fields \cite{Sneyd1994} is somehow similar to the
one obtained by Bojarevics and Romerio: it is found that cells
comprising two waves of equal frequency are always unstable.

A coupling of two modes was further studied by Davidson and Lindsay
\cite{Davidson1998} using a mechanical analog and shallow water
approximation. The obtained relation for instability is
\begin{equation}\label{eqn:beta:davidson}
\frac{J_0 B_{0,z}}{\rho_1 h_2+\rho_2 h_1} >
\frac{|\omega_x^2 - \omega_y^2|}{b_1+b_2},
\end{equation}
with $\omega$ denoting the frequency of a standing gravity wave 
with the wave vector pointing in $x$ or $y$ direction, and $b_i$ 
some coupling parameters for the different Fourier
modes. Again, square and cylindrical cells are always unstable,
because the frequency of one wave in $x$ and a second one in $y$ will be the
same. 

In some recent papers it was further pointed out that not only a magnetic
field and the wave coupling, but also the reflection at the walls is
an important ingredient for amplifying the instability
\cite{Lukyanov2001, Molokov2011}. 

Another series of articles is dedicated to the Kelvin-Helmholtz
instability in aluminium reduction cells
\cite{Moreau1986,Sneyd1992}. It is well known that the flow between
two counter-rotating liquid masses is subjected to shear layer (and
other) instabilities\cite{Moisy2004,Nore2004}. However,
the resulting waves are typically relatively short and therefore
strongly dampened by surface tension; they can not explain the
long waves observed in aluminium reduction cells. Usually, 
Kelvin-Helmholtz waves are considered less important for real cells.

Finally, MHD noise, i.e. short-wave perturbations are often observed
experimentally. They develop probably by the turbulent flow in the
cell, and are less important for the operation of the reduction cell
\cite{Bojarevics2006, Molokov2011}.

In summary, the interaction of a horizontal current with a magnetic background field $B_{0,z}$, the coupling of two waves
and the wave-reflection at the walls are considered as the crucial
elements of the sloshing instability in aluminium cells \cite{Segatz1994,Molokov2011}. There already
exist several shallow water \cite{Bojarevics1998,Zikanov2000} and full
3D models \cite{Gerbeau2006,Flueck2010} for an efficient simulation
of aluminium reduction cells.

Aluminium production is not only a good example for two phase, but
also for three phase systems. A Hoope's cell (used to refine Al) 
consists of molten pure aluminium and an aluminium
copper alloy, separated by a salt electrolyte
\cite{Dube1954,Lindsay2014}. To our knowledge, no 
interface instabilities were reported in this system -- maybe due to
the low current densities (0.35\,A/cm$^2$) and the thick electrolyte layer
($>$8\,cm). Nevertheless, the stability of interfaces in three layer
systems was already studied in two theoretical articles\cite{Sneyd1985,Zikanov2015}. 
Sneyd\cite{Sneyd1985} found the relation for instability
\begin{equation}\label{eqn:beta:sneyd}
\frac{J_0\cdot(\mu_0 J_0)}{g (\rho_3-\rho_2) } \cdot \frac{h_2\cosh(kh_2)}{\sinh^2(kh_2)} > \frac{1}{b'},
\end{equation}
with $\mu_0$, $b'$, $h_2$, $\rho_3$, $\rho_2$ and $k$ denoting the
vacuum permeability, a dimensionless value, the electrolyte layer
height, the density of bottom layer and electrolyte and the wave
number (see also figure \ref{f:dimensions}b). As he assumed the liquid
electrodes to be infinitely high, his
critical parameter does unfortunately not account for the aspect ratio
of the cell. Although the above equation looks similar to the Sele
criterion, it is not comparable: it holds only for the interaction of
a current with its own magnetic field -- there is no magnetic
background field $B_{0,z}$ present.

Zikanov was the first to investigate the sloshing instability in liquid
metal batteries \cite{Zikanov2015}. He used a mechanical analogue as
proposed by Davidson and Lindsay \cite{Davidson1998}. For the onset of
a Sele like sloshing instability, and assuming shallow layers, he found 
the condition
\begin{equation}\label{eqn:beta:zikanov}
\frac{J_0 B_{0,z}}{\rho_1 h_1} + \frac{J_0 B_{0,z}}{\rho_3
  h_3} > \left|\frac{\omega_y^2-\omega_x^2}{2}\right|,
\end{equation}
which is very similar to equation
(\ref{eqn:beta:davidson}). The thickness of the electrolyte layer does 
not appear, because Zikanov supposed it to be
very thin. Assuming thick layers with an aspect ratio in
the order of one, he finds the instability condition
\begin{equation}\label{eqn:beta:zikanov:thick}
c_1\frac{J_0 B_{0,z} l_x^2}{12\rho_1g h_1 h_2} + c_2\frac{J_0B_{0,z}l_x^2}{12\rho_3 g h_2 h_3} >\left|1-\frac{\omega_x^2}{\omega_y^2}\right| 
\end{equation}
with $c_1$ and $c_2$ denoting two geometrical constants.
He further
explored the interaction of an \emph{azimuthal} field with a
vertical current -- similar to Sneyd. He found the instability condition
\begin{equation}\label{eqn:beta:zikanov2}
\frac{\mu_0 J_0 r\cdot
  J_0}{g}\left(\frac{r^3}{48\rho_1h_2^2h_1}+\frac{r^3}{48\rho_3h_2^2h_3}+\frac{r}{16\rho_1h_2}-\frac{r}{16\rho_3h_2}\right)
  > 1
\end{equation}
with $r$ denoting the radius of a cylindrical cell.

Some of the critical parameters for onset of sloshing (equation
\ref{eqn:beta:bojarevics}, \ref{eqn:beta:davidson},  \ref{eqn:beta:zikanov} and
 \ref{eqn:beta:zikanov:thick}) predict cylindrical cells to be always
unstable. However, these criteria neglect dissipation by magnetic 
induction and viscosity as well as the influence of surface tension
\cite{Sneyd1992}. Especially induction
effects due to a flow in the cell can increase the critical values
for the onset of sloshing\cite{Descloux1994}. Further,
$B_{0,z}$ does not only destabilize the cell: at very high values it can even
suppress the instabiliy again \cite{Descloux1991}.

In this article we
will study only the influence of a vertical magnetic background field $B_{0,z}$. This
 is another step forward to understand the complex fluid dynamics 
involved in the operation of liquid metal batteries. A detailed investigation of
the interaction of currents with an azimuthal field (equations
\ref{eqn:beta:sneyd} and \ref{eqn:beta:zikanov2}) will be postponed to
future work. 

\section{Mathematical model \& implementation}
In this chapter we present a three dimensional multiphase model used for the
simulation. The main difference of our model compared to others is the way of 
computing the magnetic field. It allows simulation with very realistic 
boundary conditions. The numerical scheme is developed in close analogy
to a MHD model for one single phase \cite{Weber2013}; the multiphase
aspects are in detail explained in the literature\cite{Ubbink1997,Rusche2002,Deshpande2012}. 
Figure \ref{f:workflow} shows the general workflow. 

The flow in an incompressible, 
viscous and electrically conducting fluid of several phases
is described by the Navier-Stokes equation (NSE) \cite{Rusche2002}
\begin{equation}\label{nse}
\frac{\partial(\rho \bi u)}{\partial t} + \nabla\cdot\left(\rho\bi u\bi u\right) 
 = - \nabla p + \nabla\cdot(\rho\nu(\nabla\bi u + (\nabla\bi u)^\intercal)) 
-\rho g\bi e_z + \bi f_\text{L} + \bi f_\text{st}
\end{equation}
and $\nabla\cdot\bi u = 0$,
with $\bi u$, $t$, $\rho$, $p$, $\nu$, $g$, $\bi
f_\text{L}$ and $\bi f_\text{st}$ denoting velocity, time, density, 
pressure, kinematic viscosity, gravity, Lorentz force and surface tension, respectively. 
The unit vector $\bi e_z$ is pointing upwards. 
Introducing a modified pressure $p_\text{d}=p + \rho g z$ by adding the hydrostatic pressure, 
boundary conditions may be formulated easier and numerical errors reduced. We find
\begin{equation}
\nabla p = \nabla p_\text{d} - \rho g \bi e_z - g z\nabla\rho.
\end{equation}
Introducing the electric current density $\bi J$ and magnetic field $\bi B$, the
Lorentz force can be expressed as $\bi f_\text{L} = \bi J\times \bi B$. 
The NSE then becomes 
\begin{align}
\begin{split}
\frac{\partial(\rho \bi u)}{\partial t} + \nabla\cdot(\rho\bi u\bi u)
 = &-\nabla p_\text{d} +  g z \nabla\rho + \nabla\cdot(\rho\nu(\nabla\bi u + (\nabla\bi u)^\intercal)) 
\\&+ \bi J\times \bi B + \bi f_\text{st}.
\end{split}
\end{align}
No-slip boundary conditions are applied for velocity and an equivalent Neumann boundary condition for the pressure:
\begin{equation}
\nabla p_\text{d}=\bi J\times \bi B + \bi f_\text{st} + g z\nabla\rho.
\end{equation}
Ohms law for moving conductors 
\begin{equation}\label{eqn:ohm}
\bi J = -\sigma\nabla\phi + \sigma(\bi u\times \bi B) -\sigma\frac{\partial \bi A}{\partial t}
\end{equation}
allows for calculating the full current density in the cell. The current
density of the initial state of rest is
\begin{equation}                                                                                                                                                                                                                                                              
\bi J_0 = \frac{I}{S}\bi e_z,
\end{equation}
with $\bi A$, $\phi$, $\sigma$, $I$ and $S$ denoting
the vector potential, electric potential, electric conductivity, cell
current and cross section of the cell. We use in our model
the quasi-static approximation \cite{Herreman2015,Weber2015b,Bandaru2016}, and neglect 
the temporal derivative of the vector potential ($\partial \bi A/\partial t\approx0$).
Demanding charge conservation ($\nabla\cdot\bi J = 0$) and applying the divergence
operator to equation (\ref{eqn:ohm}), we find a Poisson equation for
the electric potential
\begin{equation}\label{mp:poisson}
\nabla\cdot(\sigma\nabla\phi)=\nabla\cdot(\sigma(\bi u\times\bi B)).
\end{equation}
As no current is flowing through the side walls of the cylinder, we apply there the boundary condition
$\nabla\phi\cdot\bi n = 0$ with $\bi n$ denoting the
surface normal vector. 
We force the perturbed current to form closed loops by adjusting the
boundary flux of $\phi$ at top and bottom according to $\bi J_0$:
\begin{equation}
\nabla\phi\cdot\bi n = -\frac{\bi J_0\cdot \bi n}{\sigma}.
\end{equation}
While not completely correct, this is a quite reasonable boundary
condition, because the current collectors often have a slightly lower
conductivity than liquid metals. 

In a last step, the magnetic field $\bi B$ is calculated using the perturbed vector
potential $\bi a$ and the magnetic field of an infinitely long cylinder 
\begin{equation}
\bi B_{0,\varphi} = \frac{\mu_0 I}{2\pi r} \bi e_\varphi
\end{equation}
as
\begin{equation}
\bi B = \nabla\times \bi a + \bi B_{0,\varphi} + \bi B_{0,z}.
\end{equation}
In the quasi-stationary limit ($\partial\bi A/\partial t\approx 0$) \cite{Bandaru2016} and using the Coulomb gauge, 
the vector potential is obtained by solving the Poisson equation
\begin{equation}
0= \frac{1}{\mu_0}\Delta\bi a + \bi J - \bi J_0
\end{equation}
with the boundary conditions obtained by Green's identity \cite{Santalo1993}
\begin{equation}\label{biotA}
\bi a(\bi r) = \frac{\mu_0}{4\pi}\int dV\frac{\bi J(\bi r')-\bi J_0(\bi r')}{|\bi r -\bi
  r'|}.
\end{equation}
This integral can be calculated much faster than Biot-Savart's law --
this is the reason why we use the vector potential.

The three different phases of the liquid metal battery are 
modelled using the volume of fluid method.
The phase fraction
$\alpha_i$ describes the fraction of fluid $i$ in a single
cell. It is determined by solving the transport equation \cite{Ubbink1997,Rusche2002}
\begin{equation}                                                                                                                                                
\frac{\partial\alpha_i}{\partial t} + \nabla\cdot\bi u \alpha_i = 0.                                                                                                                                                   
\end{equation}
All variable fluid properties are then defined by the phase fraction
as
\begin{equation}
  \rho = \sum_i \alpha_i\rho_i,\qquad
  \rho\nu  = \sum_i \alpha_i\rho_i\nu_i \quad\text{and}\quad
  \sigma = \sum_i \alpha_i\sigma_i.
\end{equation}
Note that the conductivity of the mixture can be calculated as a
serial or parallel connection of resistances\cite{Carson2005}. Both
approaches represent extreme cases. The true conductivity of a cell
depends on the angle between the interface and the current density. As
this is not known, we use the above mentioned simplified parallel circuit.
This may lead to a thinning of the electrolyte
layer and overestimate onset and short-circuit of the
cell. However, as it smoothes the conductivity at the interface, it stabilises
the simulation and is therefore the better choice.

The surface tension is modelled using the continuum surface force
(CSF) modell by Brackbill \cite{Brackbill1992,Ubbink1997,Rusche2002}, i.e. it is implemented as 
a volume force $\bi f_\text{st} = \sum_i\sum_{j\neq i}\gamma_{ij}\kappa_{ij}\delta_{ij}$, concentrated at the 
interface. The contact angle at the walls is assumed to be $90\,^\circ$.
The interface tension $\gamma_{ij}$ between phases $i$ and $j$ is assumed to be a
constant. The curvature between phase $i$ and $j$ is given as 
\begin{equation}
\kappa_{ij}\approx-\nabla\cdot\frac{\alpha_j\nabla\alpha_i - \alpha_i\nabla\alpha_j}{|\alpha_j\nabla\alpha_i - \alpha_i\nabla\alpha_j|},
\end{equation}
the Dirac delta function $\delta_{ij} = \alpha_j\nabla\alpha_i - \alpha_i\nabla\alpha_j$
ensures that the force is applied only near an interface. Finally, we find
\begin{equation}
\bi f_\text{st} \approx -\sum_i\sum_{j\neq i}\gamma_{ij} \nabla\cdot\left(\frac{\alpha_j\nabla\alpha_i - \alpha_i\nabla\alpha_j}{|\alpha_j\nabla\alpha_i - \alpha_i\nabla\alpha_j|}\right)(\alpha_j\nabla\alpha_i - \alpha_i\nabla\alpha_j).
\end{equation}
First order schemes are used for discretisation of the temporal 
derivative (Euler implicit) and the convection term (upwind); 
all other schemes are second order accurate. The grid resolution 
is 50 cells on the diameter, with a maximum aspect ratio of 3 in 
the electrolyte layer.

\section{Results}
In order to illustrate metal pad rolling in LMBs, we model
a simple cylindrical Mg$||$Sb cell of diameter $d=10\,$cm. The 
electrolyte used is NaCl-KCl-MgCl$_2$ (30-20-50 mol\%) 
\cite{Bradwell2011, Bradwell2012, Kim2013b}. If not otherwise
stated, we use the dimensions and physical properties of table
\ref{tab:physProperties}; 
the current density is $1\,$A/cm$^2$
%, the current $I=78.5\,$A
%the upper metal layer height $h_1=4.5$\,cm, the electrolyte layer
%thickness $h_E=1$\,cm, the lower metal layer height $h_3=4.5$\,cm 
and the assumed magnetic background field $B_{0,z} =10\,$mT.
This high value (approximately 200 times the magnetic field of the earth)
is chosen in order to evidence clearly the effect of metal pad rolling.
The interface tensions are 
estimated using the surface tensions as \cite{Israelachvili2011}
\begin{equation}
\gamma_{1|2} = \gamma_1 + \gamma_2 - 2.0\sqrt{\gamma_1\gamma_2}.
\end{equation}
We find $\gamma_{\text{1}|\text{2}}=0.19\,$N/m, $\gamma_{\text{1}|\text{3}}=0.016\,$N/m and $\gamma_{\text{2}|\text{3}}=0.095\,$N/m; the indices 1, 2 and 3
are refering to upper electrode, electrolyte and lower electrode, respectively. The capillary
length
\begin{equation}
\lambda = \sqrt{\frac{\gamma_{\text{1}|\text{2}}}{\Delta\rho g}} = 12\,\text{mm}
\end{equation}
is approximately 10\,\% of the cell diameter and 100\,\% of the
electrolyte thickness. We expect therefore no big influence of surface
tension on the onset of sloshing, but it may be relevant for a localised
short-circuit.

\begin{table}[t!]%\vspace{8pt}%\tiny
\centering
\caption{Physical dimensions and properties of an
  Mg$|$NaCl-KCl-MgCl$_2|$Sb cell
  \cite{Todreas2008,Kim2013b,Gale2004}.}
\begin{tabular}{lrrrrr}\hline\label{tab:physProperties}
&\multicolumn{1}{c}{h}&\multicolumn{1}{c}{$\rho$} & \multicolumn{1}{c}{$\nu$} & \multicolumn{1}{c}{$\sigma$} & \multicolumn{1}{c}{$\gamma$}\\
%\cmidrule(lr){2-2}\cmidrule(lr){3-3}\cmidrule(lr){4-4}\cmidrule(lr){5-5}\cmidrule(lr){6-6}
\hline
&\multicolumn{1}{c}{cm}&\multicolumn{1}{c}{kg/m$^3$}& \multicolumn{1}{c}{ m$^2$/s}& \multicolumn{1}{c}{S/m}&\multicolumn{1}{c}{N/m}\\
\midrule
top layer (1)      & 4.5 & 1577 & $6.7\cdot10^{-7}$ & $3.62\cdot 10^6$ & 0.54\\
electrolyte (2) & 1   & 1715 & $6.8\cdot10^{-7}$ & 80 & 0.09\\
bottom layer (3)    & 4.5 & 6270 & $1.96\cdot10^{-7}$ & $8.66\cdot10^5$  & 0.37\\
\hline\end{tabular}\vspace{3pt}\end{table}
In the initial state of the system, both interfaces are flat; gravity,
hydrostatic
pressure, an axial magnetic field $\bi B_{0,z}$ and a current $I$ from bottom
to top are applied.

\subsection{Driving mechanism of the instability}\label{ch:instability_mechanism}
\begin{figure}[t!]
\centering
\includegraphics[width=\textwidth]{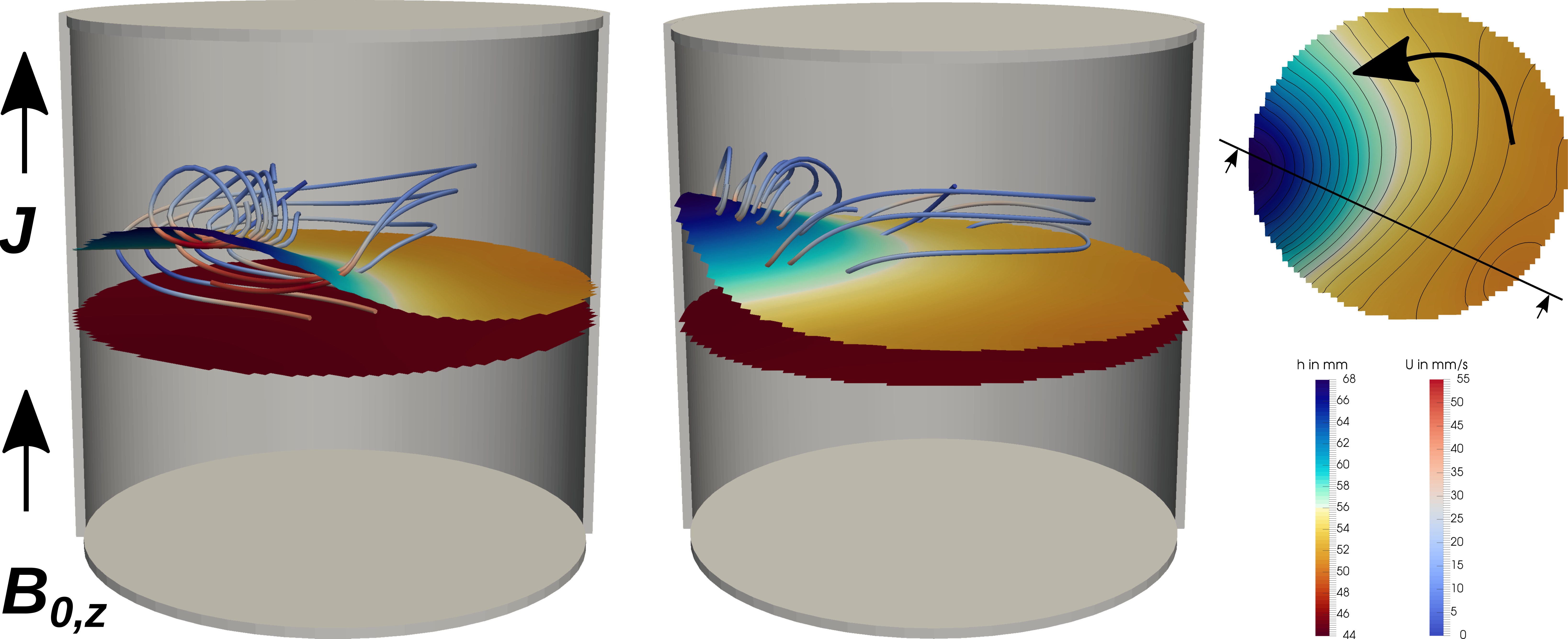}
\caption{Streamlines of velocity, surface elevation and direction of
  rotation for a saturated sloshing instability in a liquid metal
  battery ($I=120$\,A, $B_z=10$\,mT, $h_1=h_3=4.5$\,cm, $h_2=1$\,cm).}
\label{f:flow}
\end{figure}
Figure \ref{f:flow} illustrates a typical saturated sloshing
instability in an LMB. The blue crest of electrolyte is localy
concentrated, has steep flanks and rotates anticlockwise. The trough
is vast, flat and smooth. Crest and trough are not symmetric: the
crest tries to catch up the trough. The principal flow matches well 
to the deplacement of the wave crest. The streamlines in the upper
metal close directly above the crest, or in a long azimuthal flow.

\begin{figure}[b!]
\centering
 \subfigure[]{\includegraphics[height=0.28\textwidth]{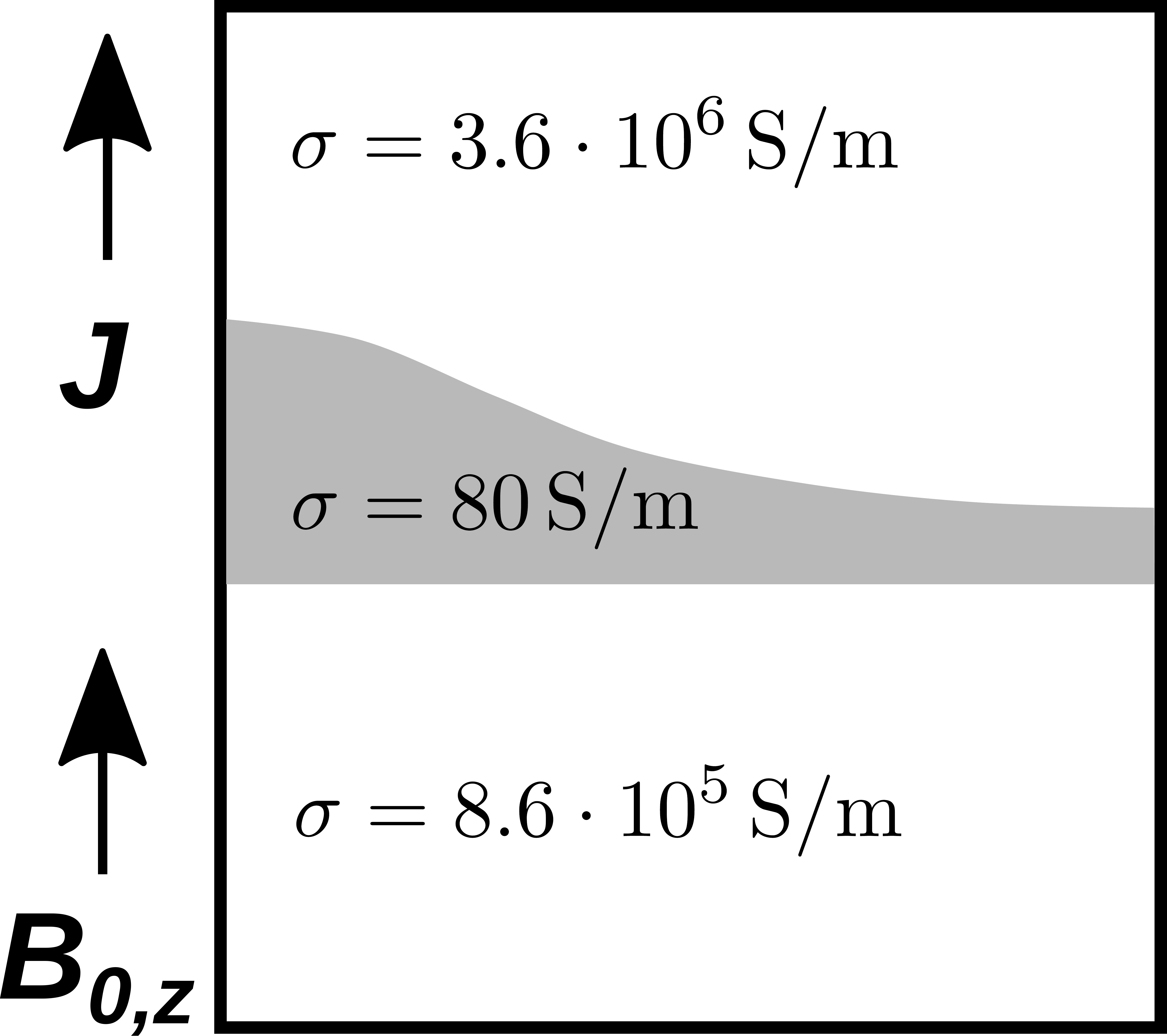}}\hfill
 \subfigure[]{\includegraphics[height=0.28\textwidth]{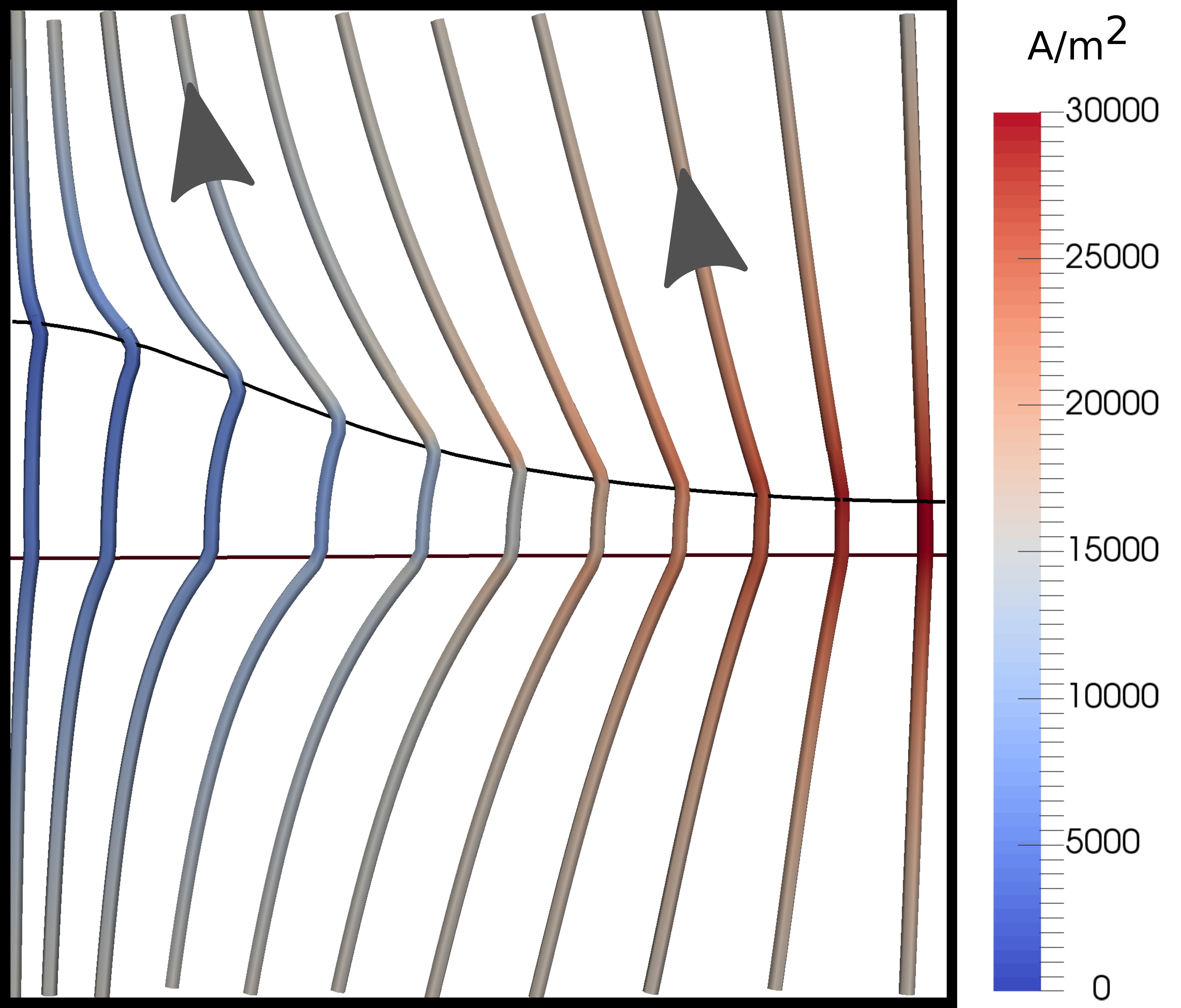}}\hfill
 \subfigure[]{\includegraphics[height=0.28\textwidth]{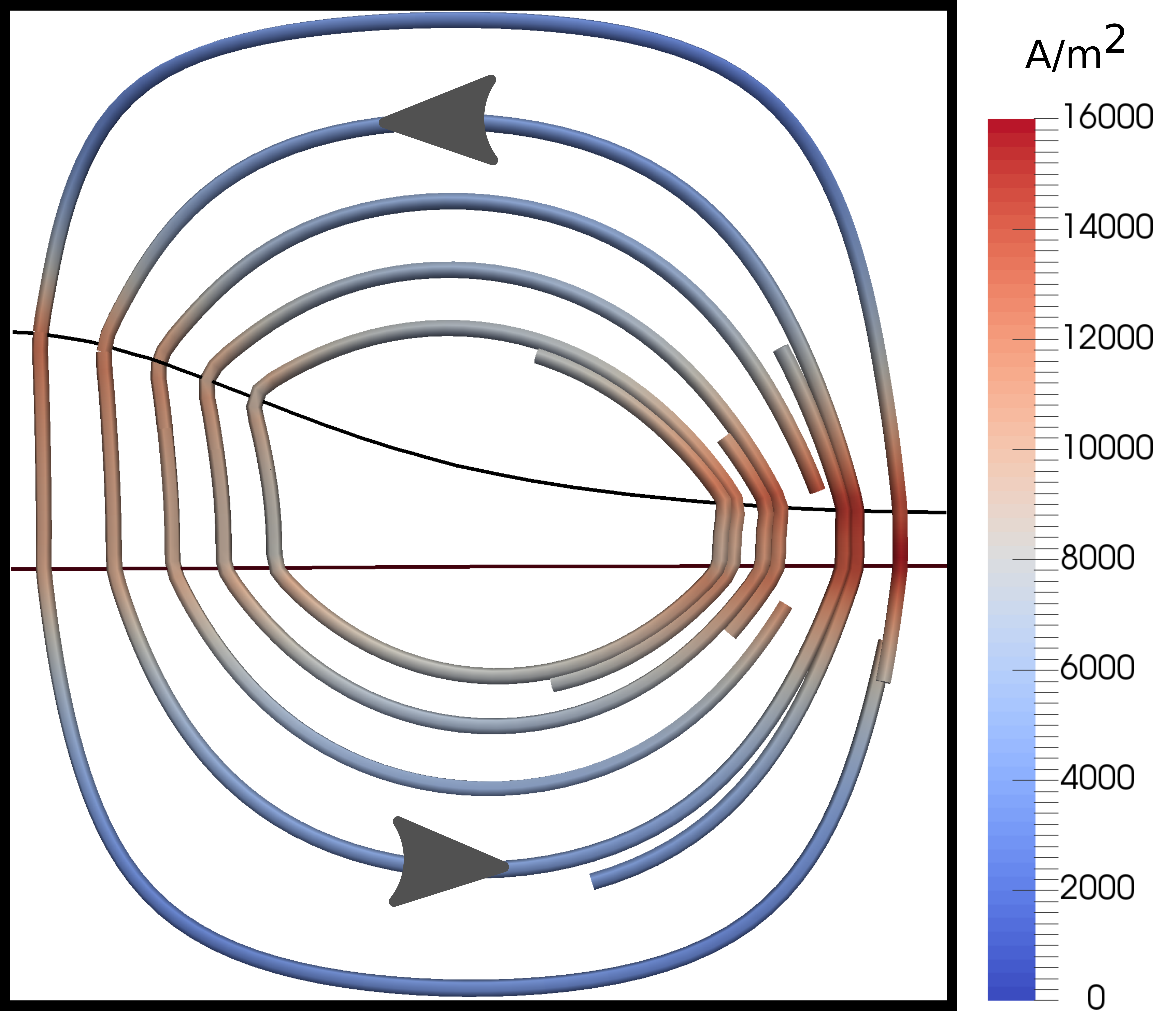}}\hfill
\caption{Conductivity (a), complete current $\bi J$ (b) and compensation or
  perturbation current $\bi j$
(c) for a deformed interface between upper metal and electrolyte layer
($I=120$\,A, $B_z=10$\,mT, $h_1=h_3=4.5$\,cm, $h_2=1$\,cm). The
location of the plane is indicated in figure \ref{f:flow}.}
\label{f:2dcurrent}
\end{figure}

In order to explain the instability mechanism, we plot several
features on a plane indicated by the black line in figure
\ref{f:flow}. The cell current $\bi J$ concentrates on the pinched side of the
electrolyte (figure \ref{f:2dcurrent}b). Substracting the current
of an unperturbed cell ($\bi J_0$), we obtain the compensation current
$\bi j$ in figure 
\ref{f:2dcurrent}c. This current distribution is not surprising; it
strongly resembles the simple model described in the
introduction. Please note that we force the perturbed current to
close within the cell by applying Neumann boundary conditions for the
electric potential.

Applying a constant vertical magnetic field $\bi B_{0,z}$ and denoting the
unperturbed magnetic field as $\bi B_{0,\varphi}$ and the perturbed one as $\bi b$, the
Lorentz force can be expressed by four relevant summands:
\begin{equation}
\bi f_\text{L} = \bi J_0\times\bi b + \bi
j\times \bi B_{0,\varphi} + \bi j\times \bi B_{0,z} + \bi j\times\bi b.
\end{equation}
Please note that $\bi J_0\times\bi B_{0,\varphi}$ is a pure gradient and drives
no flow; $\bi B_{0,z}$ is parallel to $\bi J_0$.
\begin{figure}[t!]
\centering
 \subfigure[\,$\bi J_0\times\bi b$:
 stabilising]{\includegraphics[height=0.34\textwidth]{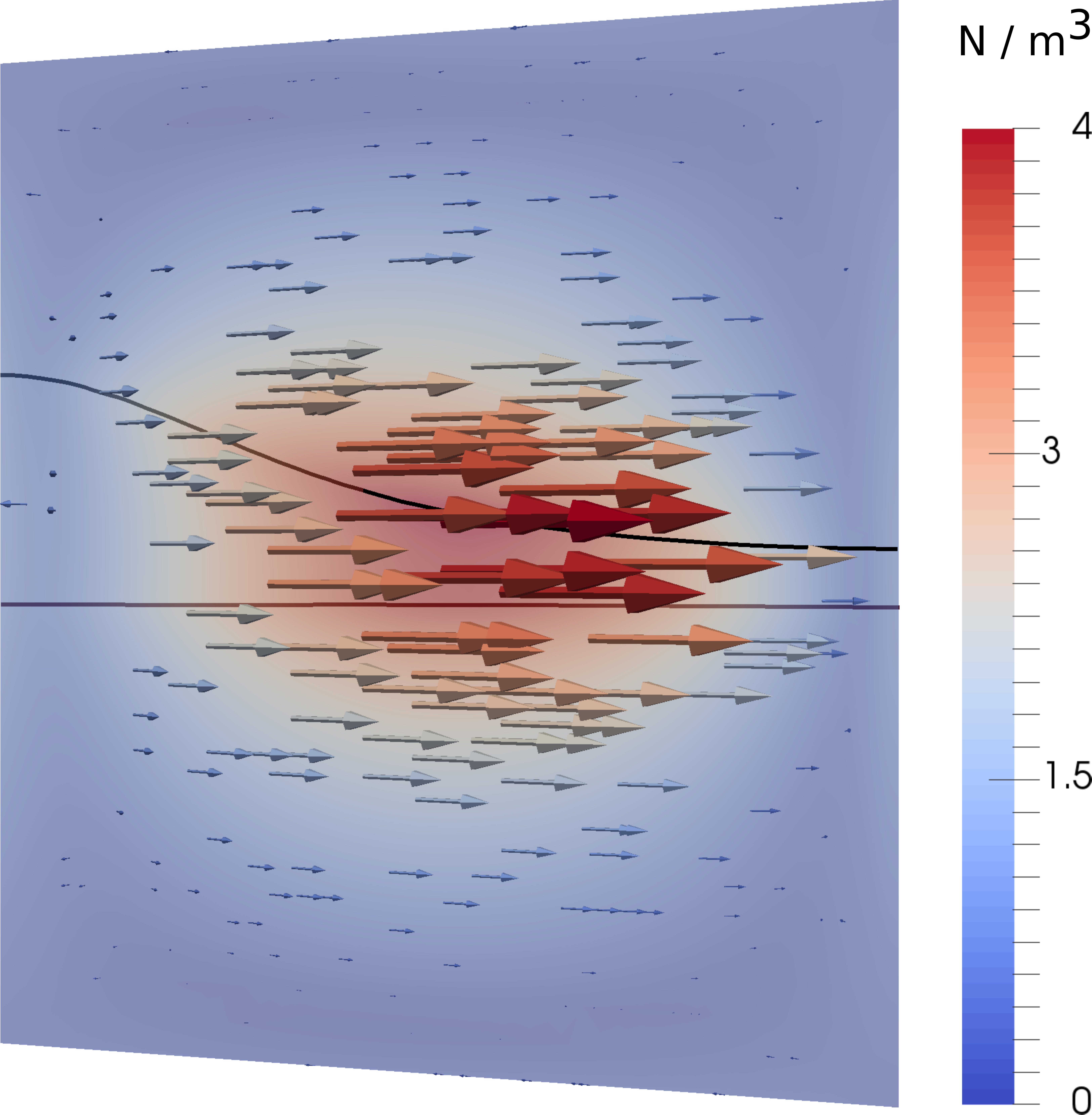}}\hfill
 \subfigure[\,$\bi j\times\bi b$:
 stabilising]{\includegraphics[height=0.34\textwidth]{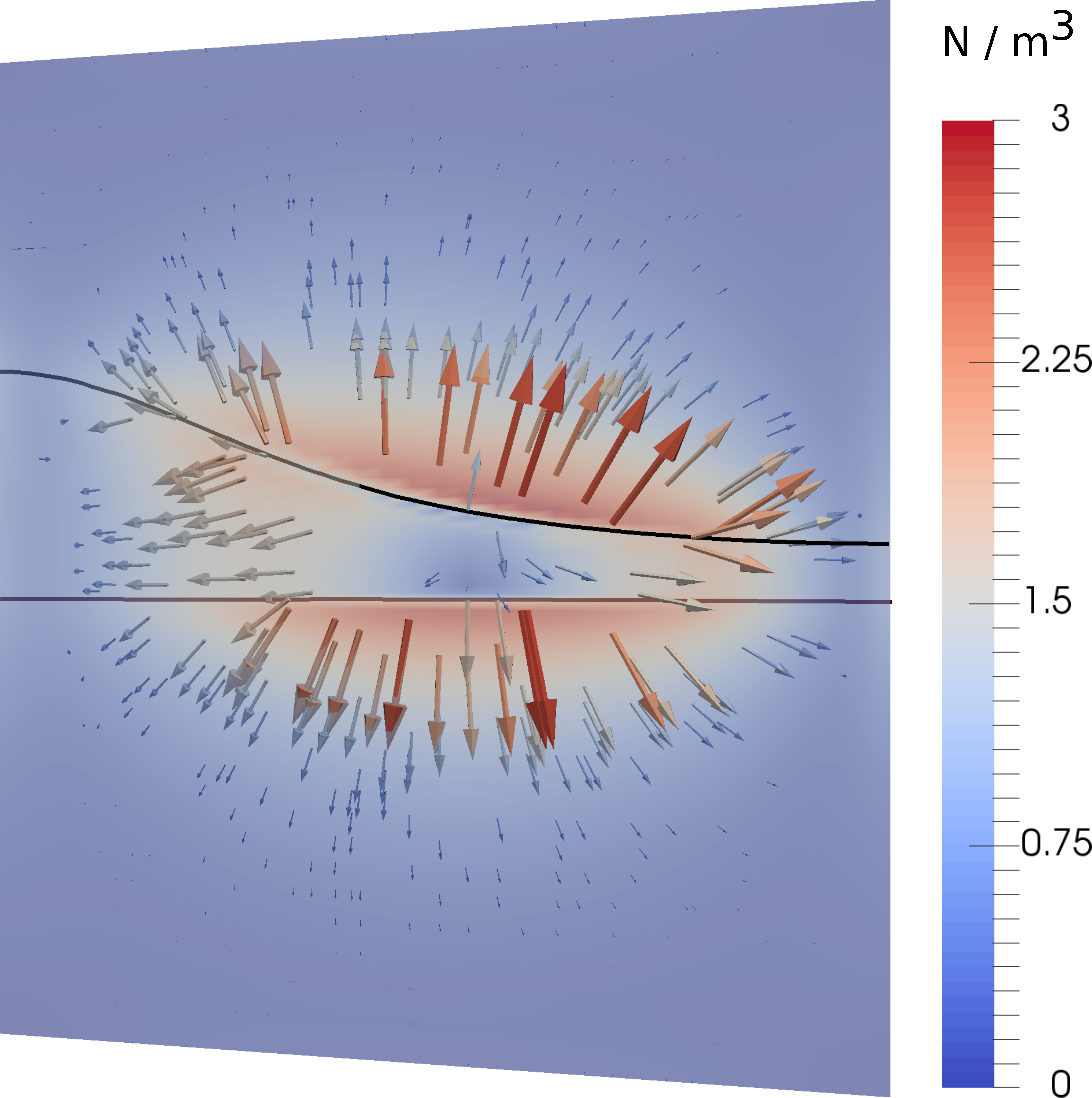}}\hfill
 \subfigure[\,$\bi j\times\bi B_{0,\varphi}$:
 destabilising]{\includegraphics[height=0.34\textwidth]{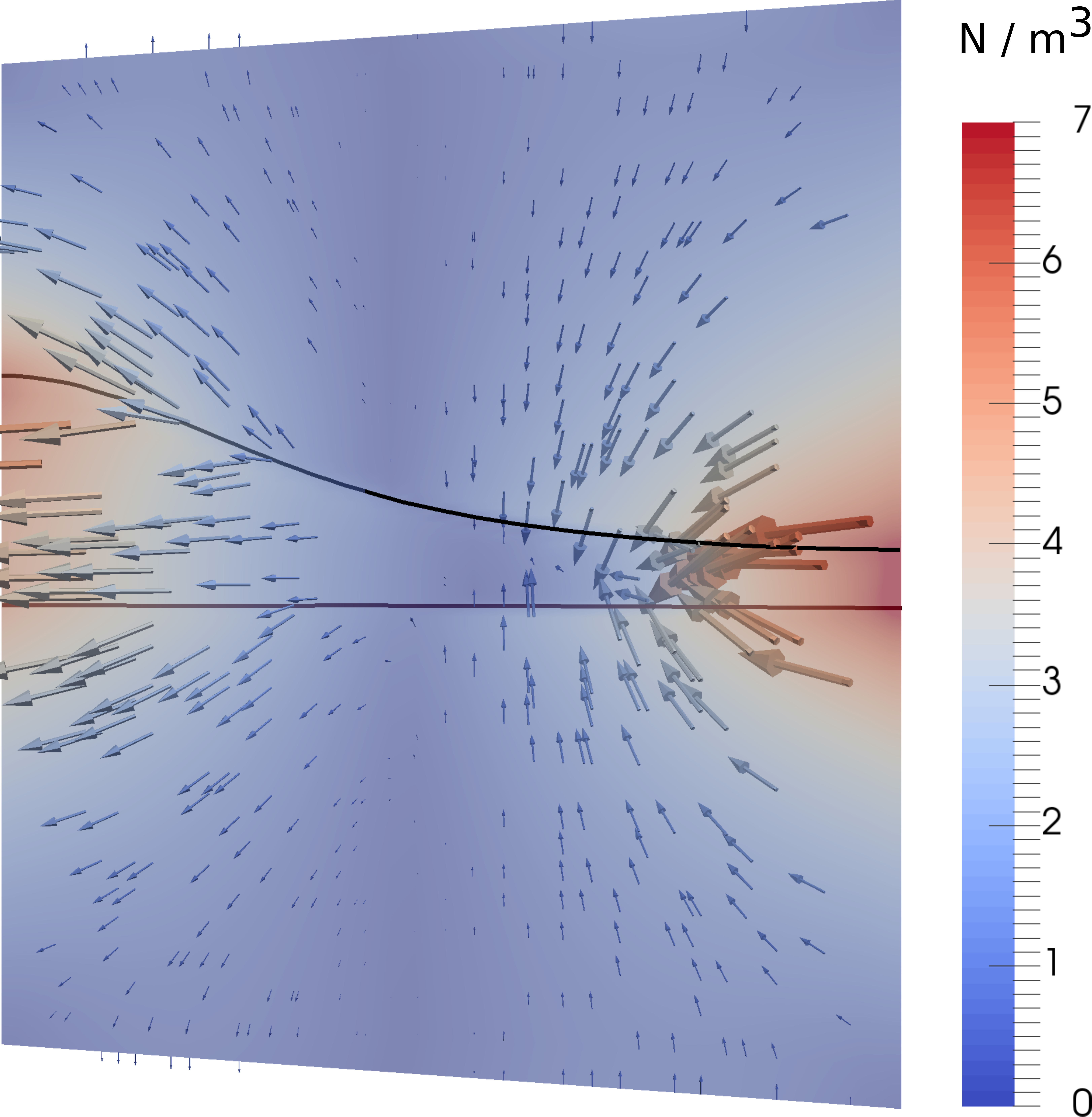}}\hfill
 \subfigure[\,$\bi j\times \bi B_{0,z}$:
 destabilising]{\includegraphics[height=0.34\textwidth]{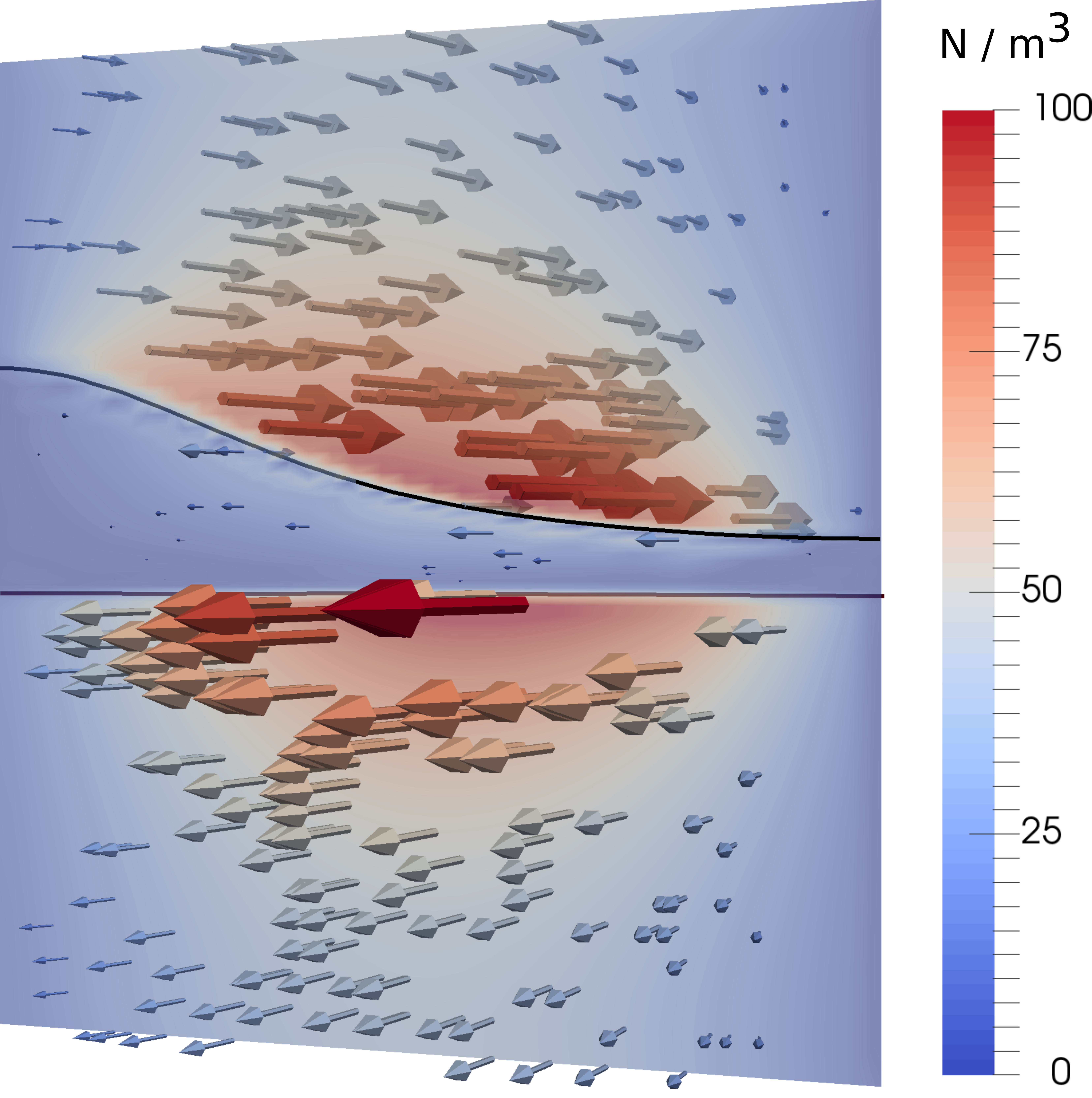}}\hfill
\caption{Different Lorentz forces for a sloshing instability. The
  prescribed current of $I=120$\,A flows upwards, the background
  magnetic field $B_{0,z}=10$\,mT points upwards ($h_1=h_3=4.5$\,cm,
  $h_2=1$\,cm). The location of the plane is indicated in figure
  \ref{f:flow}.}
\label{f:force2d}
\end{figure}

In figure \ref{f:force2d} we illustrate the four relevant force components in
the same plane as before. Firstly, we notice that the perturbed magnetic field
$\bi b$ always stabilises the interface (figure \ref{f:force2d}a and
b). However, the interaction of a horizontal current and the azimuthal
field ($\bi j\times \bi B_{0,\varphi}$) destabilises the electrolyte (figure
\ref{f:force2d}c). This was already suggested by Zikanov
\cite{Zikanov2015}, but without considering the two damping
forces. 

Finally, we show in figure \ref{f:force2d}d the interaction of a
horizontal current and $\bi B_{0,z}$. This force is by far the largest and
considered as the main source of the sloshing instability
\cite{Molokov2011}.  We observe almost no force in the electrolyte; the 
upper metal is driven anti-clockwise, the lower clockwise and both forces 
are equally large. Considering that $\bi B_{0,z}$ points upwards, the
observed Lorentz force can already be deduced easily from the current
distribution in figure \ref{f:2dcurrent}c. The force in the upper metal
will lead to a rotating flow (figure \ref{f:flow}), pushing the crest of electrolyte in 
front of it. This may explain the rotation of the wave, but not the
increase of its amplitude. Indeed, a reflection of
the sloshing fluid at the cylinder wall will lead to a deformation of the
upper metal-electrolyte interface \cite{Lukyanov2001}. This coupling at 
the walls is assumed to be essential for the instability
\cite{Molokov2011}.

\subsection{Characterisation of the instability and parameter studies}
\begin{figure}[b!]
\centering
 \subfigure[]{\includegraphics[width=0.5\textwidth]{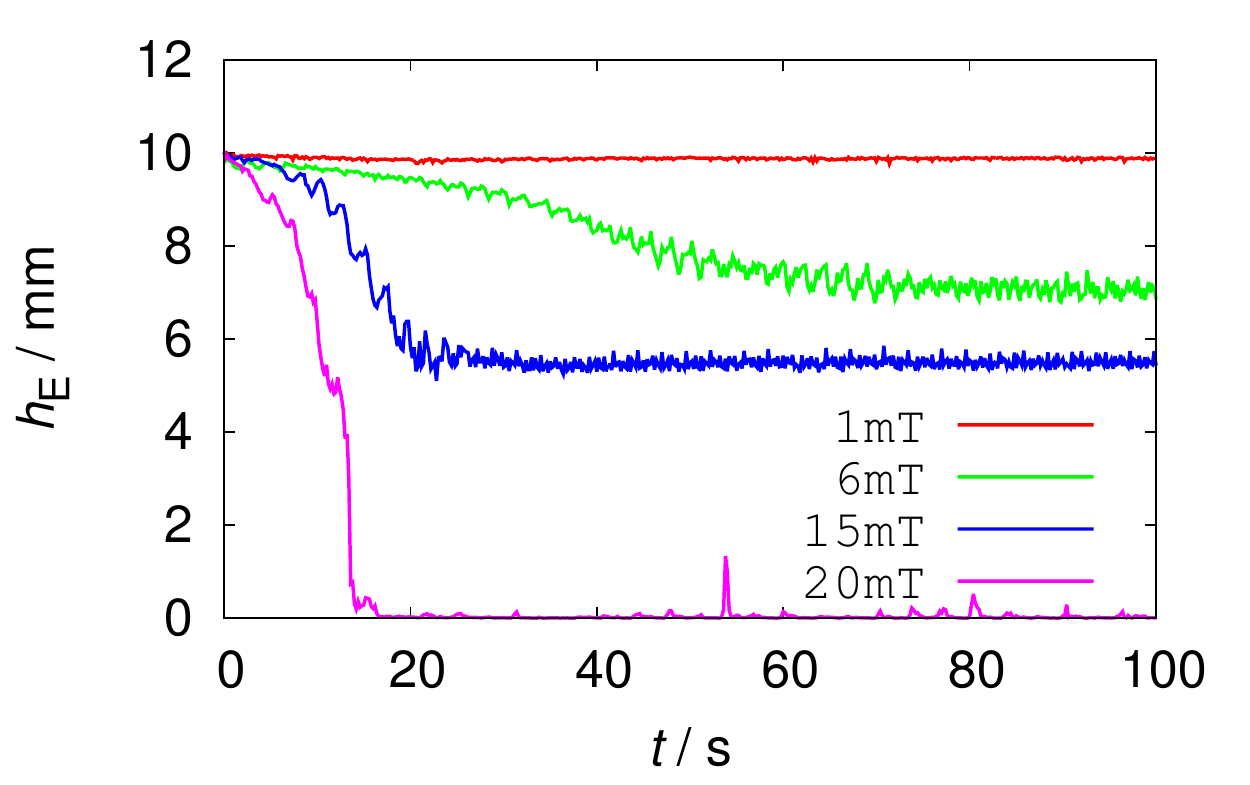}}\hfill
 \subfigure[]{\includegraphics[width=0.5\textwidth]{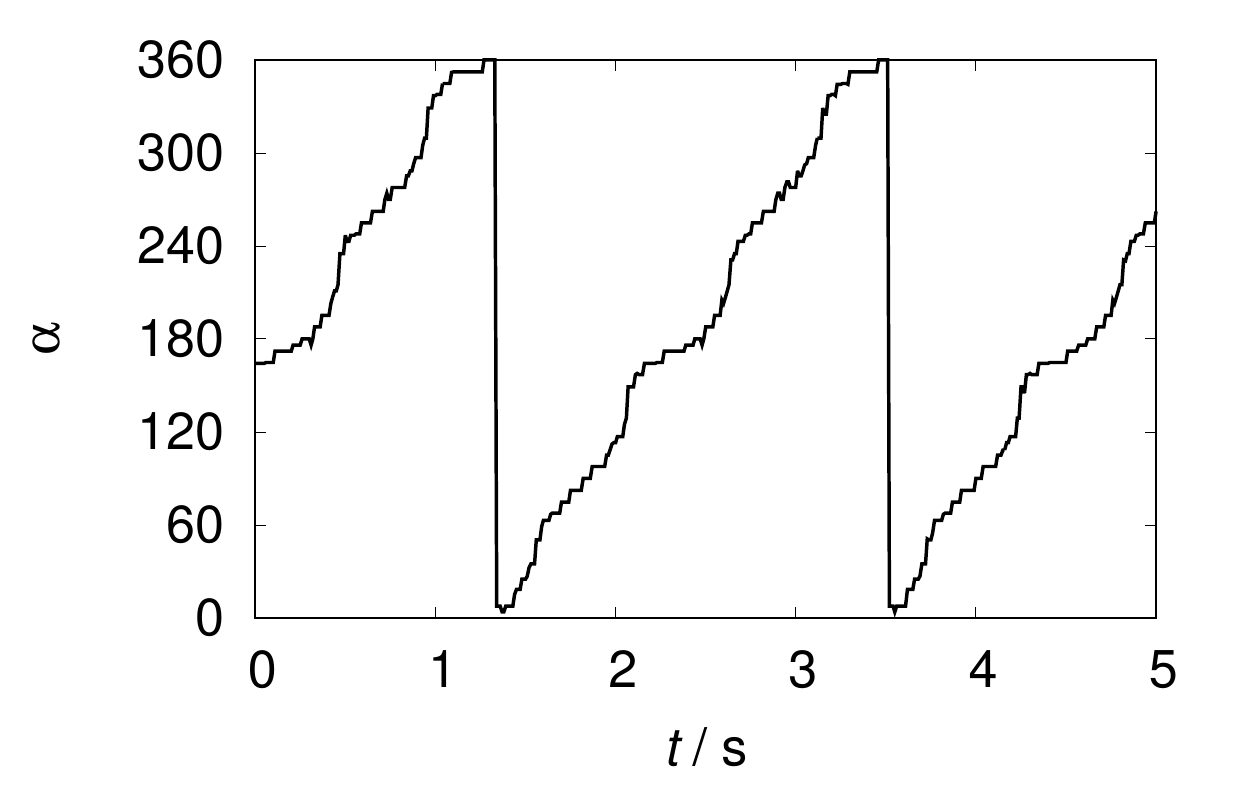}}
\caption{Temporal evolution of the minimal salt layer thickness
for different background magnetic fields (a), and of the corresponding angle
(b) for $B_{0,z}=6$\,mT ($I=78.5$\,A, $h_1=h_3=4.5$\,cm, $h_2=1$\,cm).}
\label{f:tracking}
\end{figure}
In this section we will study the influence of several parameters
on the sloshing instability in liquid metal batteries. We simulate the
same Mg$||$Sb cell as in the preceeding case and vary the cell current
$I$, the background field $B_{0,z}$, the heights of the top layer $h_1$, electrolyte $h_2$
and bottom metal $h_3$, as well as the electrolyte conductivity
$\sigma_2$, the top metal density $\rho_1$ and the viscosity
$\nu$. Note that a change of a layer thickness implies also a change
of the cells aspect ratio. We start our simulation with plane
interfaces and track the minimal salt layer thickness $h_E$ over time until
reaching a saturated state (see figure \ref{f:tracking}a). The period
of rolling (figure \ref{f:tracking}b) is determined, too. It is expected 
to deviate only slightly from the period of the gravity wave
\cite{Pedchenko2009,Molokov2011}. Further we expect that the natural
sloshing frequencies in LMBs can be suitably approximated by wave
solutions of two-layer systems due to the high density of the bottom
alloy. To check this presumption we deduce dispersion relations of the
two-phase and the three-phase system for cylindrical vessels using
potential theory. For the angular frequency $\omega$ in a two-phase
system we receive
\begin{align}
&\omega_{mn}^{2} = \frac{(\rho_2 - \rho_1)g\frac{\epsilon_{mn}}{R} +
  \gamma_{1|2}\left(\frac{\epsilon_{mn}}{R}
  \right)^3}{\rho_1\coth(\frac{\epsilon_{mn}}{R}h_1) +
  \rho_2\coth(\frac{\epsilon_{mn}}{R}h_2)} 
\label{eq:2}
\end{align}
where $R$ and $g$ denote the radius and gravitational
acceleration. The wave number $\epsilon_{mn}$ corresponds to the
$n$th roots of the first derivative $J_{m}^{'}(\epsilon_{mn}) = 0$ of
the $m$th-order Bessel function of the first kind with valid modes $m
= 0,1,2,\dots$ and $n=1,2,\dots$, respectively. Within a three-layer
system natural frequencies can be expressed by
\begin{align}
&\omega_{1|2,mn}^{2} = \frac{(\rho_2 - \rho_1)g\frac{\epsilon_{mn}}{R} +
  \gamma_{1|2}\left(\frac{\epsilon_{mn}}{R}
  \right)^3}{\rho_1\coth(\frac{\epsilon_{mn}}{R}h_1) +
  \rho_2\left(\coth(\frac{\epsilon_{mn}}{R}h_2) +
  \frac{A_{mn}^{2}}{A_{mn}^{1}}\frac{1}{\sinh(\frac{\epsilon_{mn}}{R}h_2)}\right)}\label{eq:12}
\\
&\omega_{2|3,mn}^{2} = \frac{(\rho_3 - \rho_2)g\frac{\epsilon_{mn}}{R} +
  \gamma_{2|3}\left(\frac{\epsilon_{mn}}{R}
  \right)^3}{\rho_3\coth(\frac{\epsilon_{mn}}{R}h_3) +
  \rho_2\left(\coth(\frac{\epsilon_{mn}}{R}h_2) +
  \frac{A_{mn}^{1}}{A_{mn}^{2}}\frac{1}{\sinh(\frac{\epsilon_{mn}}{R}h_2)}\right)}
\label{eq:23}
\end{align}
with $A_{mn}^{1}$ and $A_{mn}^{2}$ denoting the amplitudes of the
upper and lower interface. The deviation of the frequency
$\omega_{12mn}^{2}$ of the upper surface, which is  expected to be
mainly excited by the sloshing instability, from the two-layer frequency
is only manifested in the term 
\begin{align}
\label{eq:dev}
\frac{A_{mn}^{2}}{A_{mn}^{1}}\frac{1}{\sinh(\frac{\epsilon_{mn}}{R}h_2)}.
\end{align} 
Hence, expression (\ref{eq:dev}) can be exploited to analyse the
validity of the two-layer approximation. From there, the two-layer
relation is suitable if, e.g., the lower amplitude becomes small enough,
$A_{mn}^{2} \ll 1$, or if the aspect ratio of the electrolyte becomes
sufficiently large $h_2 / R \gg 1$. In order to be able to calculate
the three-layer frequencies the amplitude ratios in both relations
(\ref{eq:12}) and (\ref{eq:23}) must be eliminated yielding the secular equation
\begin{align}
&a\omega^4 + b\omega^2 - c = 0 \nonumber \\
\text{with}\nonumber \\
&a = \frac{\rho_{2}^{2}}{\sinh(\frac{\epsilon_{mn}}{R}h_2)^2} - \left(
  \rho_2 \coth(\frac{\epsilon_{mn}}{R}h_2)+\rho_3
  \coth(\frac{\epsilon_{mn}}{R}h_3)\right)  \nonumber \\ 
& \ \, \times \left( \rho_2 \coth(\frac{\epsilon_{mn}}{R}h_2) + \rho_1
  \coth(\frac{\epsilon_{mn}}{R}h_1)\right)  \nonumber \\ 
&b = \left((\rho_2 -\rho_1)g \frac{\epsilon_{mn}}{R} ++
  \gamma_{1|2}\left(\frac{\epsilon_{mn}}{R} \right)^3
  \right)\left(\rho_2 \coth(\frac{\epsilon_{mn}}{R}h_2)+\rho_3
  \coth(\frac{\epsilon_{mn}}{R}h_3)\right) \nonumber \\ 
& \ \, + \left((\rho_3 - \rho_2)g \frac{\epsilon_{mn}}{R} +
  \gamma_{2|3}\left(\frac{\epsilon_{mn}}{R}
  \right)^3\right)\left(\rho_2
  \coth(\frac{\epsilon_{mn}}{R}h_2)+\rho_1
  \coth(\frac{\epsilon_{mn}}{R}h_1)\right) \nonumber \\ 
&c = \left((\rho_2 -\rho_1)g \frac{\epsilon_{mn}}{R} +
  \gamma_{1|2}\left(\frac{\epsilon_{mn}}{R} \right)^3 \right)
  \left((\rho_3 - \rho_2)g \frac{\epsilon_{mn}}{R} +
  \gamma_{2|3}\left(\frac{\epsilon_{mn}}{R} \right)^3\right). 
\end{align}
For the first mode that has the wave number $\epsilon_{11} = 1.841$ we find the period $T_{1|2}
= 2.09$\,s with the two-layer formula and $T_{1|2} = 2.11$\,s 
and $T_{2|3} = 0.45$\,s with the three-layer
formula. The values for the upper interface ($T_{1|2}$) are almost
equal for both formulas; the value obtained in the simulation of 
our standard case is $T_{1|2} = 2.18$\,s.
\begin{figure}[t!]
\centering
\includegraphics[width=0.7\textwidth]{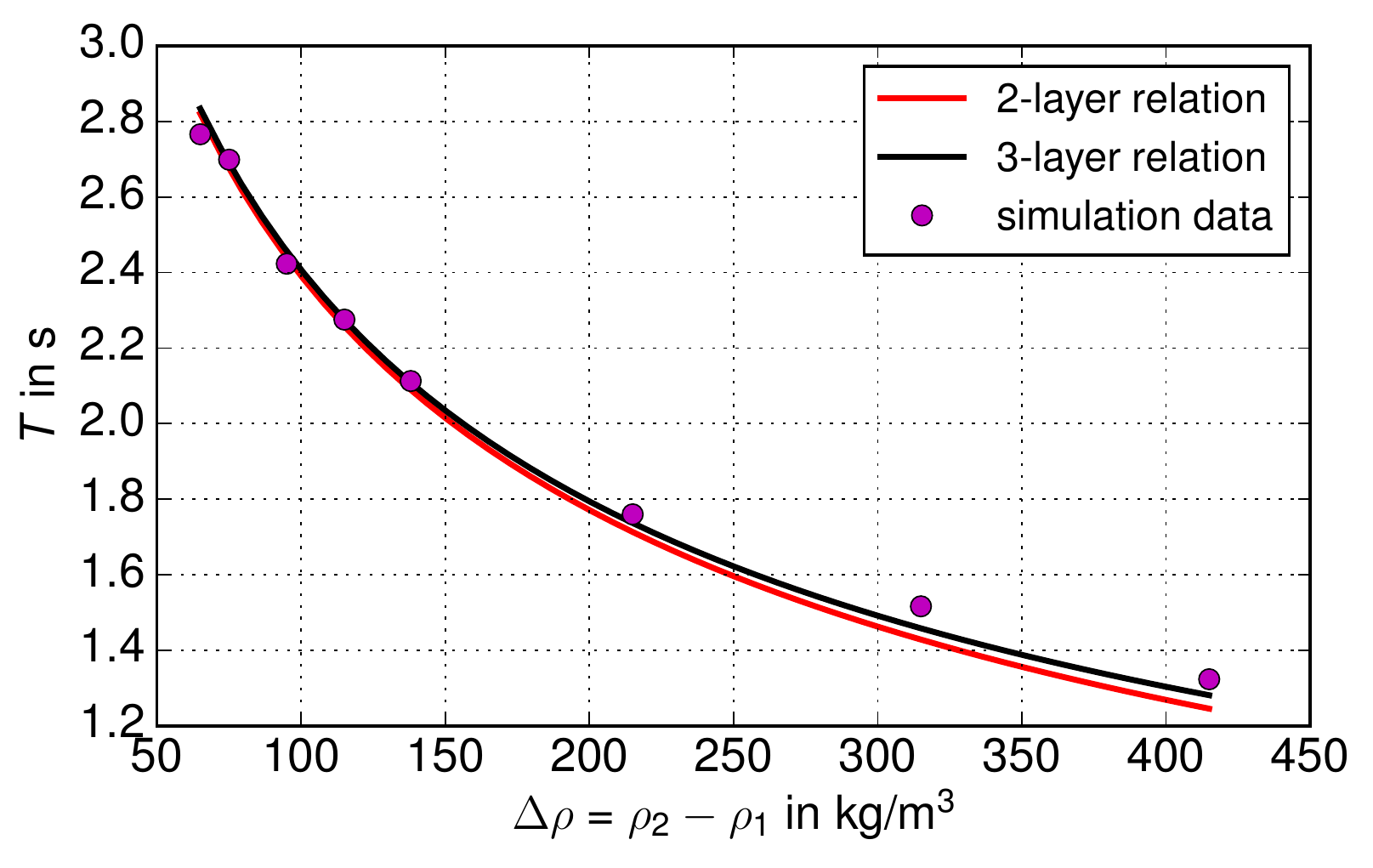}
\caption{Theoretical and simulated period of metal pad rolling with
changing density $\rho_1$ ($I=78.5$\,A,
$B_z=10$\,mT, $h_1=h_3=4.5$\,cm, $h_2=1$\,cm, $\rho_2=1715$\,kg/m$^3$,
$\rho_3=6270$\,kg/m$^3$).}
\label{f:T}
\end{figure}
Figure \ref{f:T} shows the periods for various simulations with
changing density compared to both the theoretical 2-layer (\ref{eq:12}) 
and 3-layer (\ref{eq:23}) formula. While both agree suitably with the 
numerical data, the three-layer dispersion relation matches slightly 
better for the large density differences. All in all, the two-layer relation 
has been confirmed as a suitable approximation for our liquid battery model. 
Nevertheless, the dynamics of the lower interface play a certain role for the 
evolution of short-circuits, as outlined in section \ref{sec:short-cicuit}, 
and therefore may not be fully neglected.

\begin{figure}[htb!]
\centering
 \subfigure[]{\includegraphics[width=0.45\textwidth]{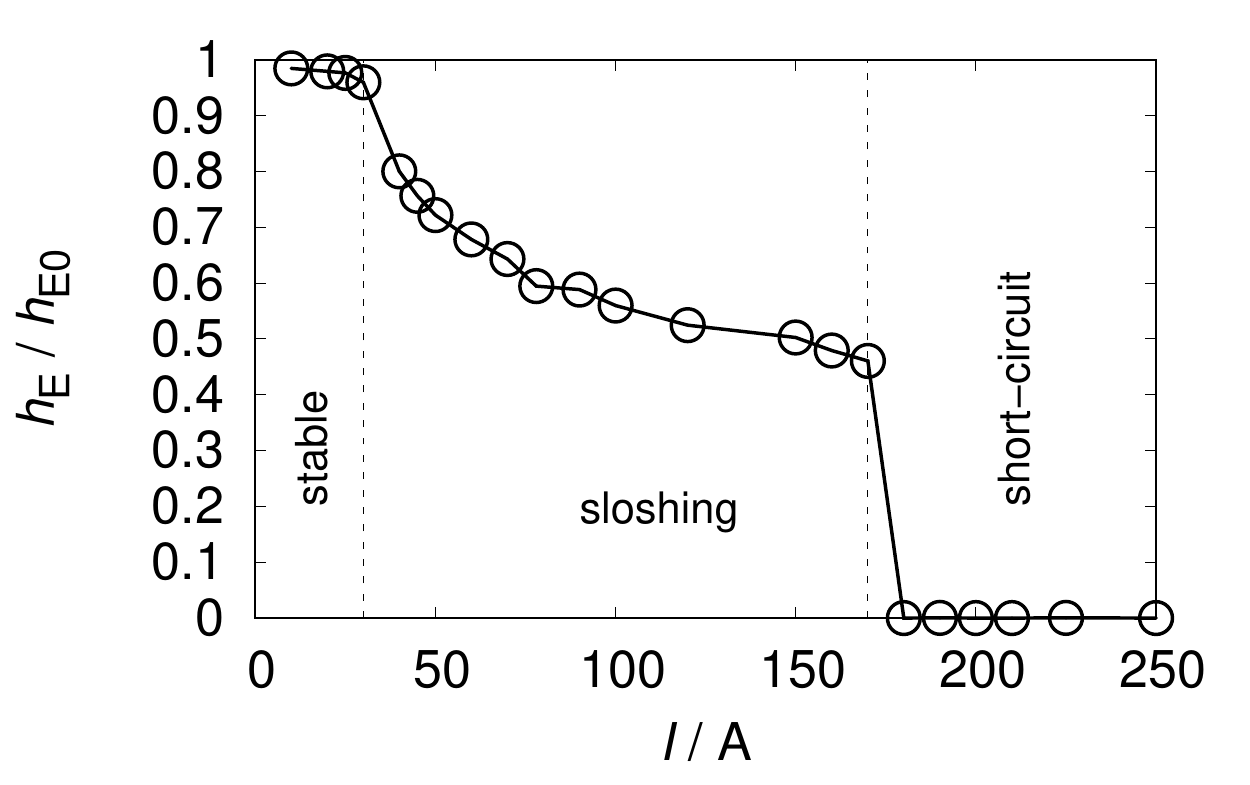}}\hfill
 \subfigure[]{\includegraphics[width=0.45\textwidth]{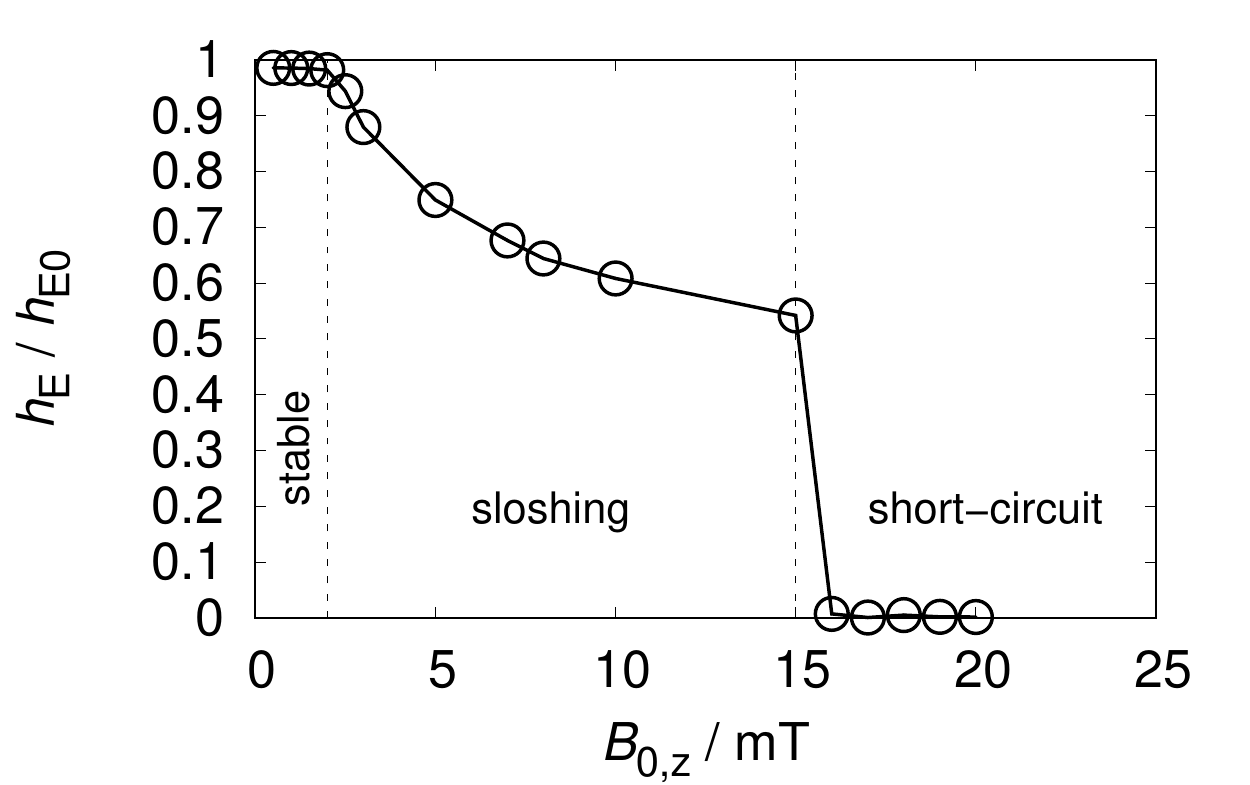}}\hfill
 \subfigure[]{\includegraphics[width=0.45\textwidth]{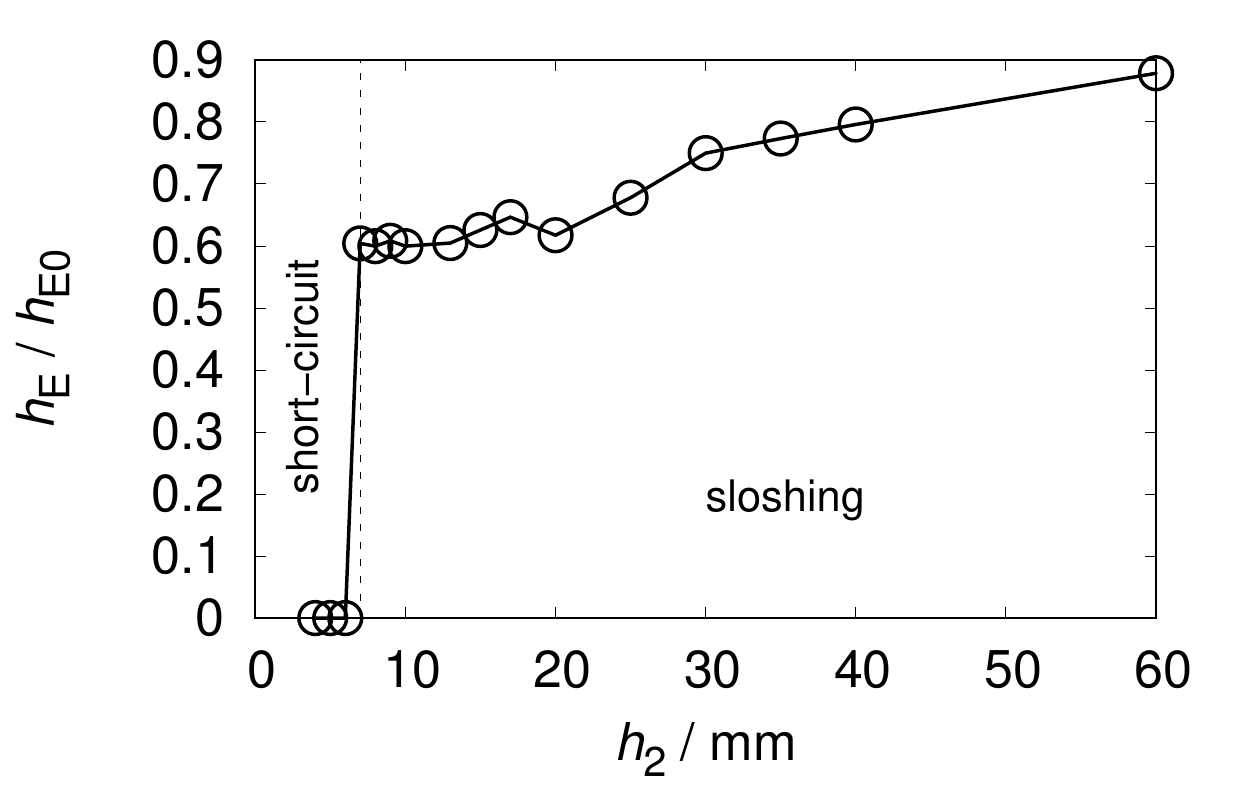}}\hfill
 \subfigure[]{\includegraphics[width=0.45\textwidth]{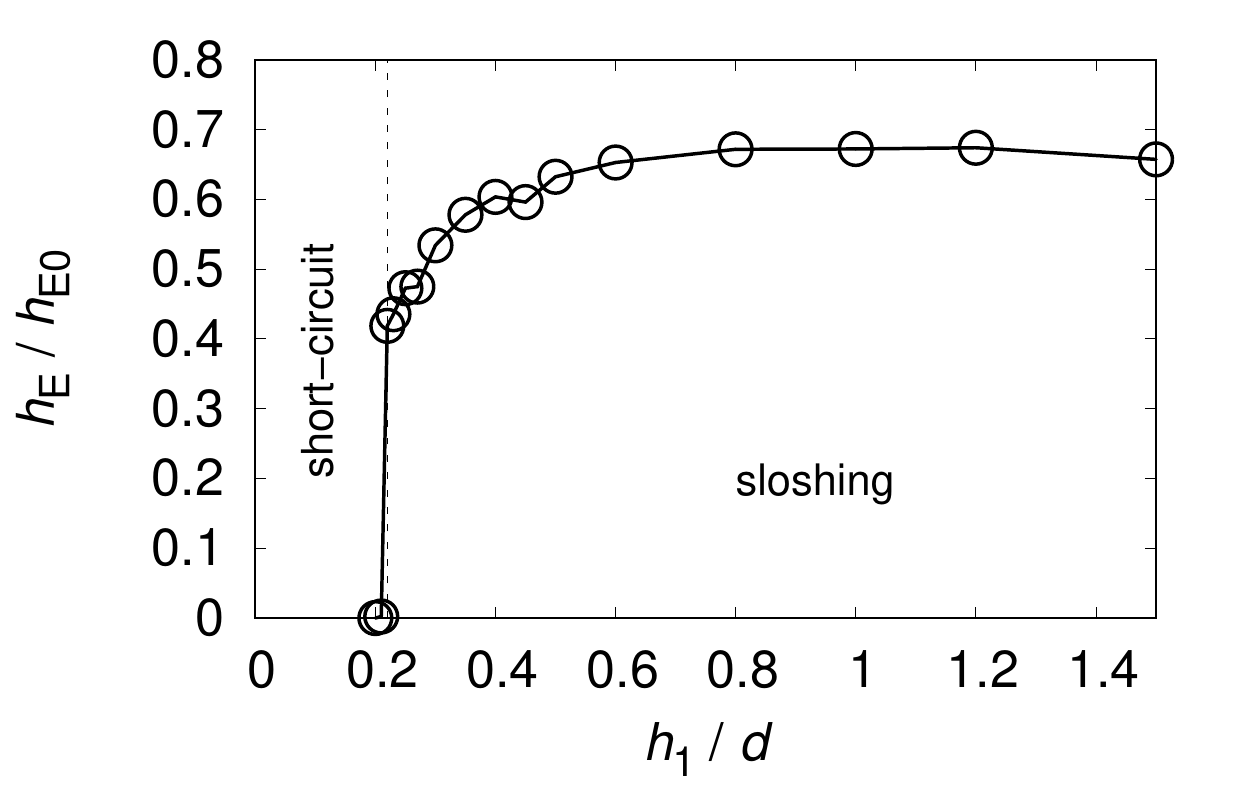}}\hfill
 \subfigure[]{\includegraphics[width=0.45\textwidth]{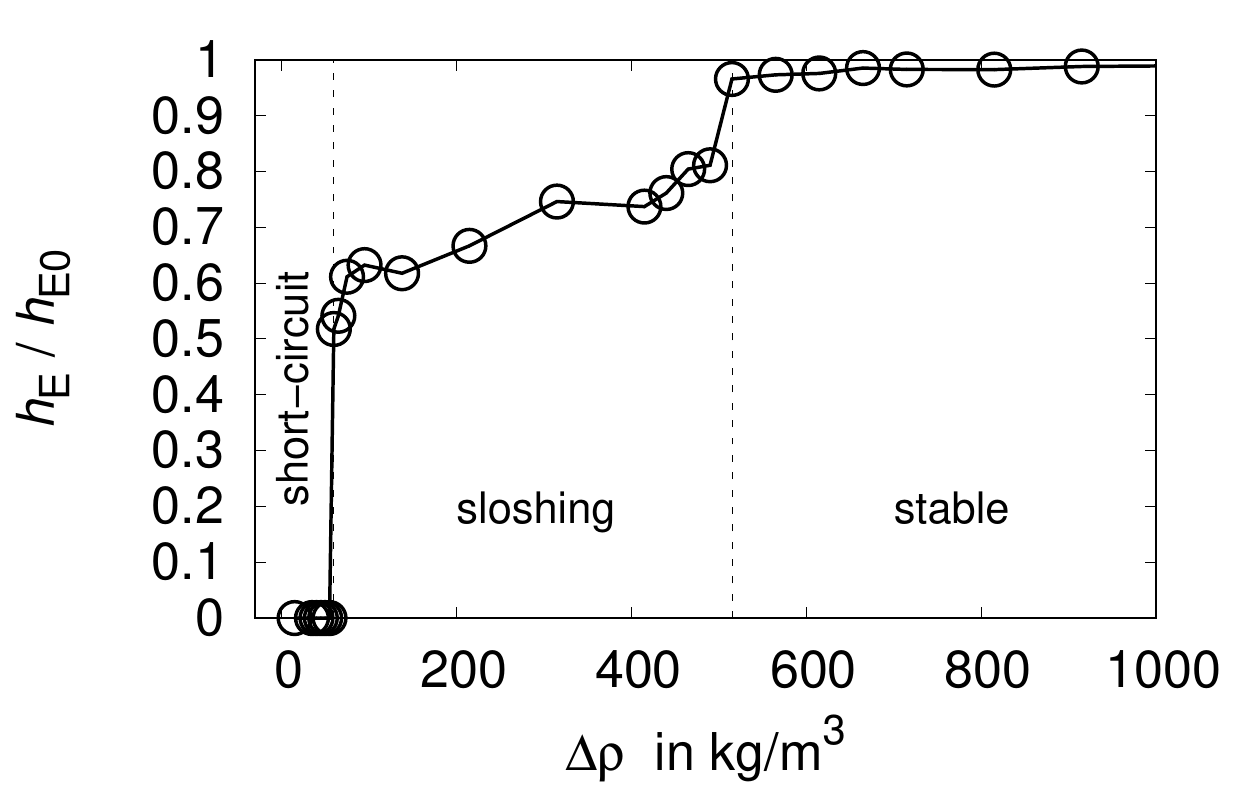}}\hfill
 \subfigure[]{\includegraphics[width=0.45\textwidth]{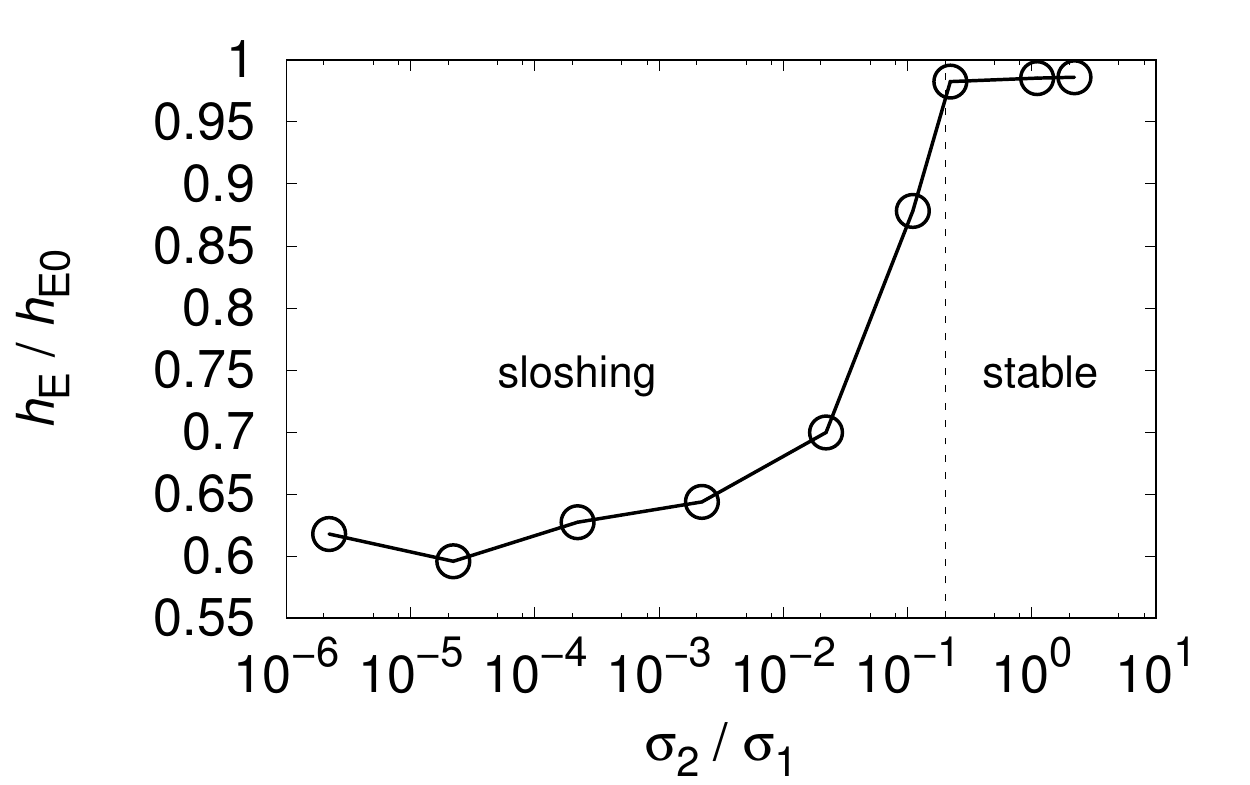}}\hfill
\caption{Minimal relative height of the salt layer depending on the 
cell current (a), the vertical magnetic background (b), the initial
heights of the salt (c) and the upper metal layer (d), the density difference
between upper metal and electrolyte (e) and the ratio of salt to upper 
metal conductivity (f). If not
being the variable quantity, the following values are used:
$I=78.5$\,A, $B_{0,z}=10$\,mT, $h_1=h_3=4.5$\,cm, $h_2=1$\,cm. For the
material parameters, see table \ref{tab:physProperties}.}
\label{f:para}
\end{figure}

In a first step of our parameter study we characterize the influence
of cell current and magnetic field: both are amplifying the
instability. Figure \ref{f:para}a shows the height of the salt layer
$h_E$ (divided by its initial height $h_{E0}$) depending 
on the cell current. Until 30\,A we do not observe any deformation of
the interface, the cell is stable. Later on, the electrolyte's minimal height
decreases with the current. This behaviour represents a bifurcation. At 170\,A we observe a sudden rupture of
the interface. Changing the magnetic background field (figure 
\ref{f:para}b) gives a very similar result -- with bifurcation points
at $2$ and $15$\,mT. 

In a second step we examine the influence of the initial heights
of the upper metal and the electrolyte layer. A shallow electrode and/or electrolyte
is more unstable -- see figure \ref{f:para}c and d. A fully stable cell 
 can not be observed; even very high layers suffer some
deformation. However, the short-circuit again appears very suddenly
when reaching an electrolyte layer thickness of $7$\,mm or an aspect
ratio of the top layer of $0.2$. In real cells, the electrolyte layer is
supposed to be between 4 and 10\,mm thick; the height of the upper electrode 
depends on the required capacity of the cell.

In a third step we explore the influence of density and electric 
conductivity. In figure \ref{f:para}e we show the minimal salt layer
thickness depending on the density difference between top metal and
salt $\Delta\rho$. As for current and magnetic field, a 
bifurcation appears at $\Delta\rho = 515\,$kg/m$^3$. The cell
fails suddenly below of $\Delta\rho = 60\,$kg/m$^3$. Note that 
the density gap between electrolyte and
bottom metal is
always very high -- we never observed a considerable deformation of
the lower interface. Decreasing the ratio of electric conductivity
(figure \ref{f:para}f) only one bifurcation occurs while no
short-circuit can be observed. Real LMBs usually have a conductivity
ratio of 10$^{-5}$. As the interface deformations does not change much
for a conductivity ratio between 10$^{-6}$ and 10$^{-3}$, it might be
possible to simulate with a higher conductivity of the salt layer in
order to avoid numerical problems.

\begin{figure}[t!]
\centering
\includegraphics[width=0.5\textwidth]{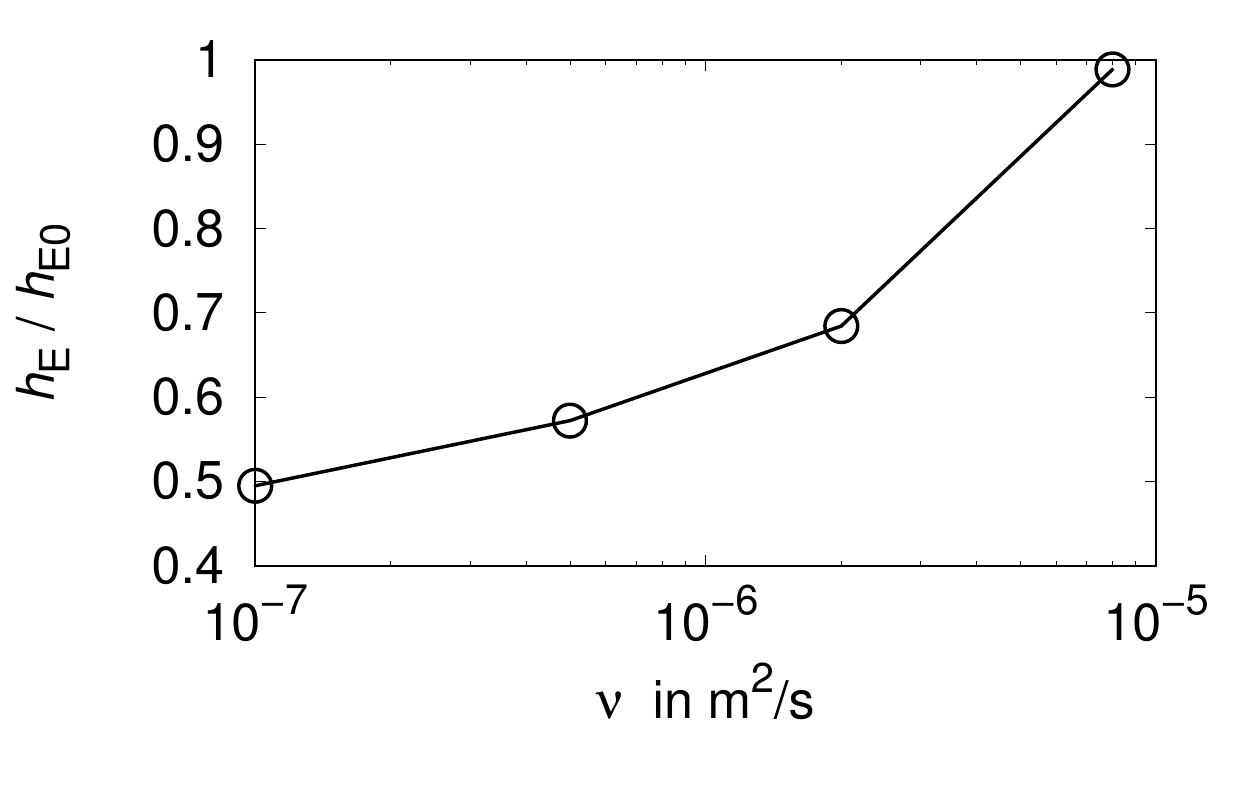}
\caption{Minimal relative height of the salt layer for changing
  viscosity of the fluids. The same viscosity is assumed for all
  phases ($I=78.5$\,A, $B_z=10$\,mT, $h_1=h_3=4.5$\,cm, $h_2=1$\,cm).}
\label{f:viscosity}
\end{figure}

Finally, we study the influence of viscosity, assuming the same viscosity 
for all phases. Typical viscosities in the order
of ($10^{-7} \dots 10^{-6}$\,m$^2$/s) considerably dampen the instability (figure
\ref{f:viscosity}). Viscosity should therefore be included in a
dimensionless number describing the sloshing in LMBs. 

In order to compare better the influence of the various parameters on
metal pad rolling, we use equations (\ref{eqn:beta:sele}),
(\ref{eqn:beta:davidson}), (\ref{eqn:beta:zikanov}) and (\ref{eqn:beta:zikanov:thick}) 
to define four different dimensionless parameters:
\begin{align}
\beta_\text{Sele} &= \frac{I B_{0,z}}{g(\rho_2-\rho_1) h_1h_2}\\[1em]
\beta_\text{Davidson} &= \frac{J_0 B_{0,z} d}{g(\rho_1 h_2+\rho_2 h_1)}\\[1em]
\beta_\text{Zikanov, thin layers}  &= \frac{J_0 B_{0,z} d}{g \rho_1 h_1} + \frac{J_0 B_{0,z} d}{g \rho_3   h_3}\\[1em]
\beta_\text{Zikanov, thick layers} &= \frac{J_0 B_{0,z} d^2}{12g\rho_1 h_1 h_2} + \frac{J_0B_{0,z}d^2}{12g\rho_3 h_2 h_3}
\end{align}
In its original meaning, these parameters describe only the onset, but
not the nonlinear part of the instability. While the first and second
one were developed for only two phases, the last two should work for
three phases of a real LMB. All but the last parameters were developed 
using the shallow water approximation, i.e. for shallow layers; it is 
therefore not straightforward to apply them to our cell.
\begin{figure}[t!]
\centering
 \includegraphics[width=0.8\textwidth]{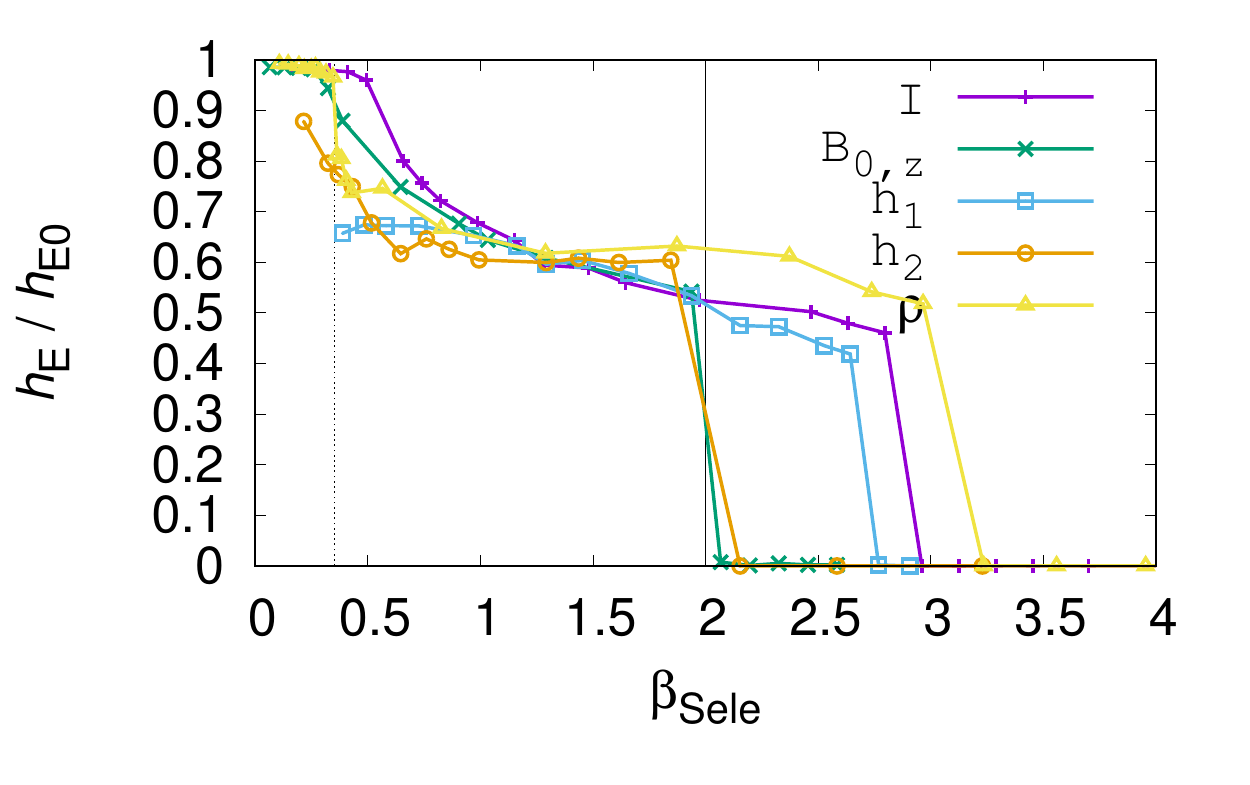}
\caption{Remaining minimal salt layer thickness depending on
the dimensionless parameter $\beta$ as defined by Sele
\cite{Sele1977,Molokov2011}. 
The value of the cell 
current $I$, the vertical field $B_{0,z}$, the height of upper metal
and electrolyte layer $h_1$ and $h_2$ as well as the density
are varied.}
\label{f:beta:sele}
\end{figure}

\begin{figure}[h!]
\centering
 \subfigure[]{\includegraphics[width=0.5\textwidth]{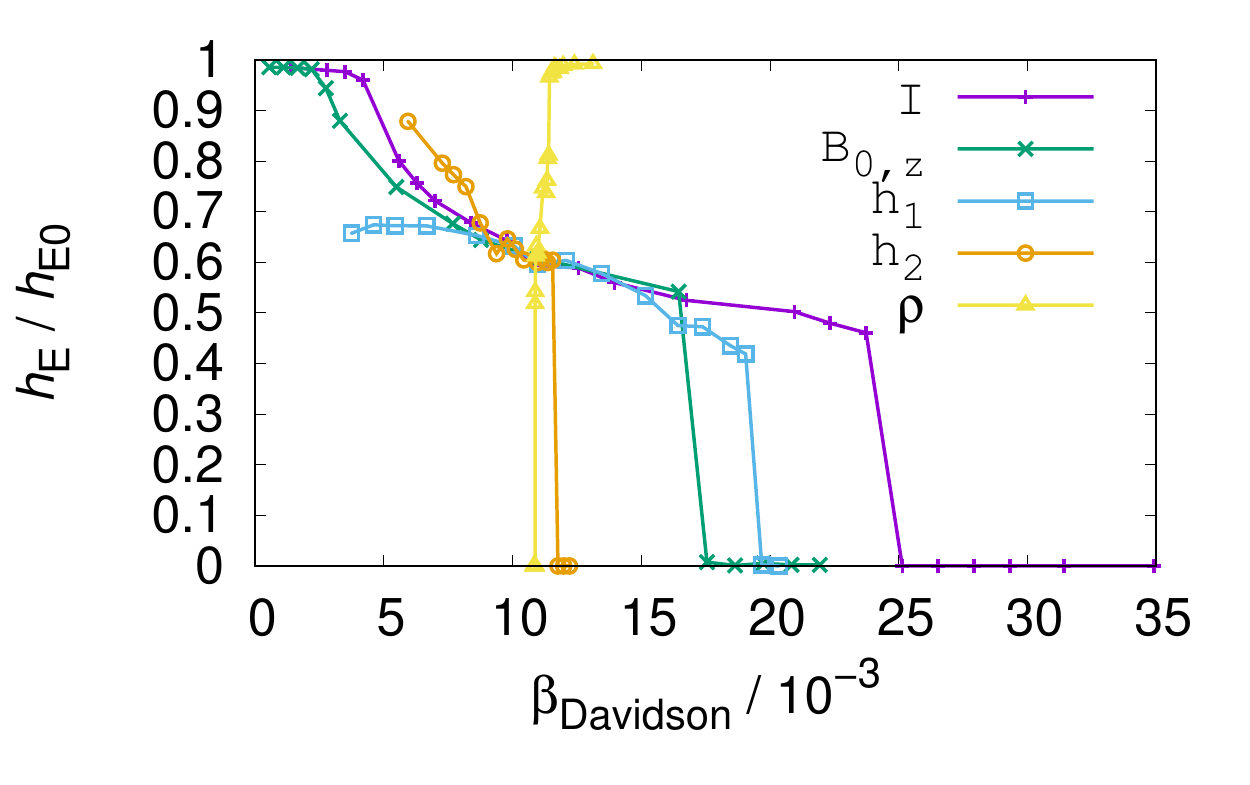}}\hfill
 \subfigure[]{\includegraphics[width=0.5\textwidth]{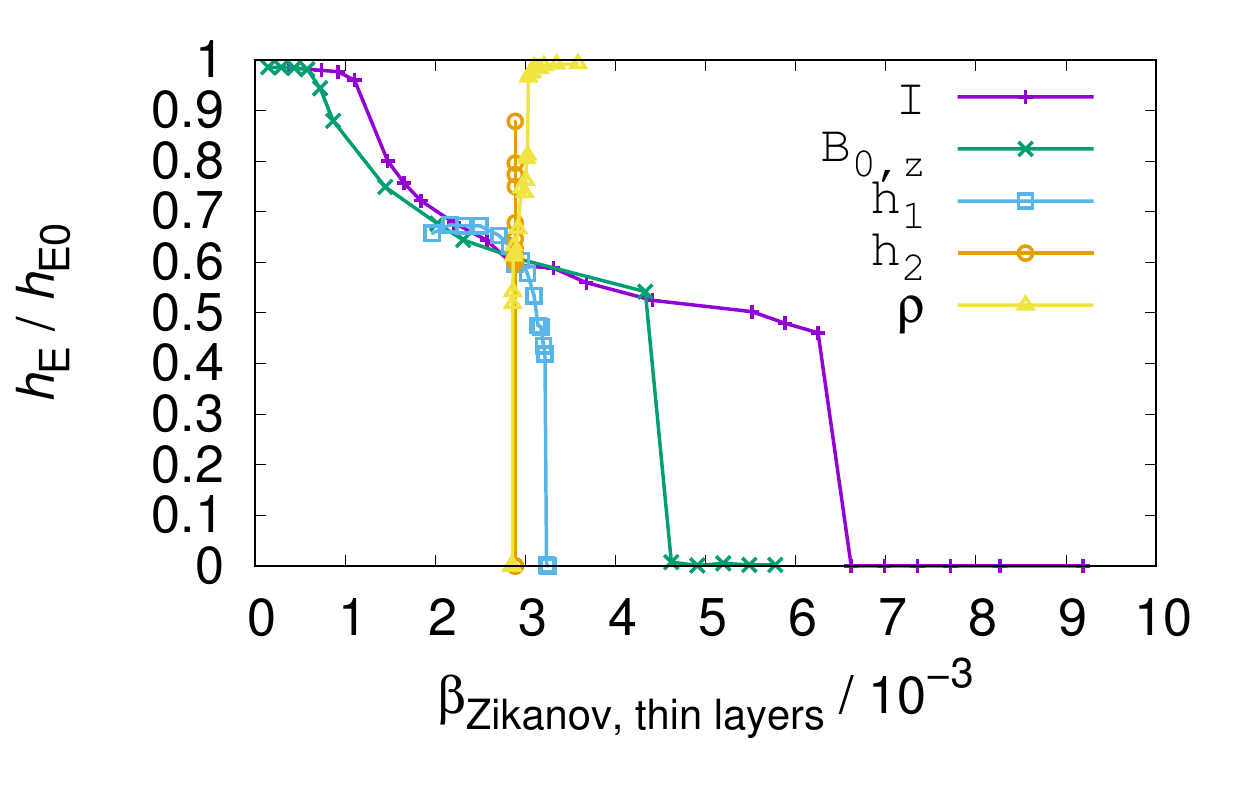}}\hfill
 \subfigure[]{\includegraphics[width=0.5\textwidth]{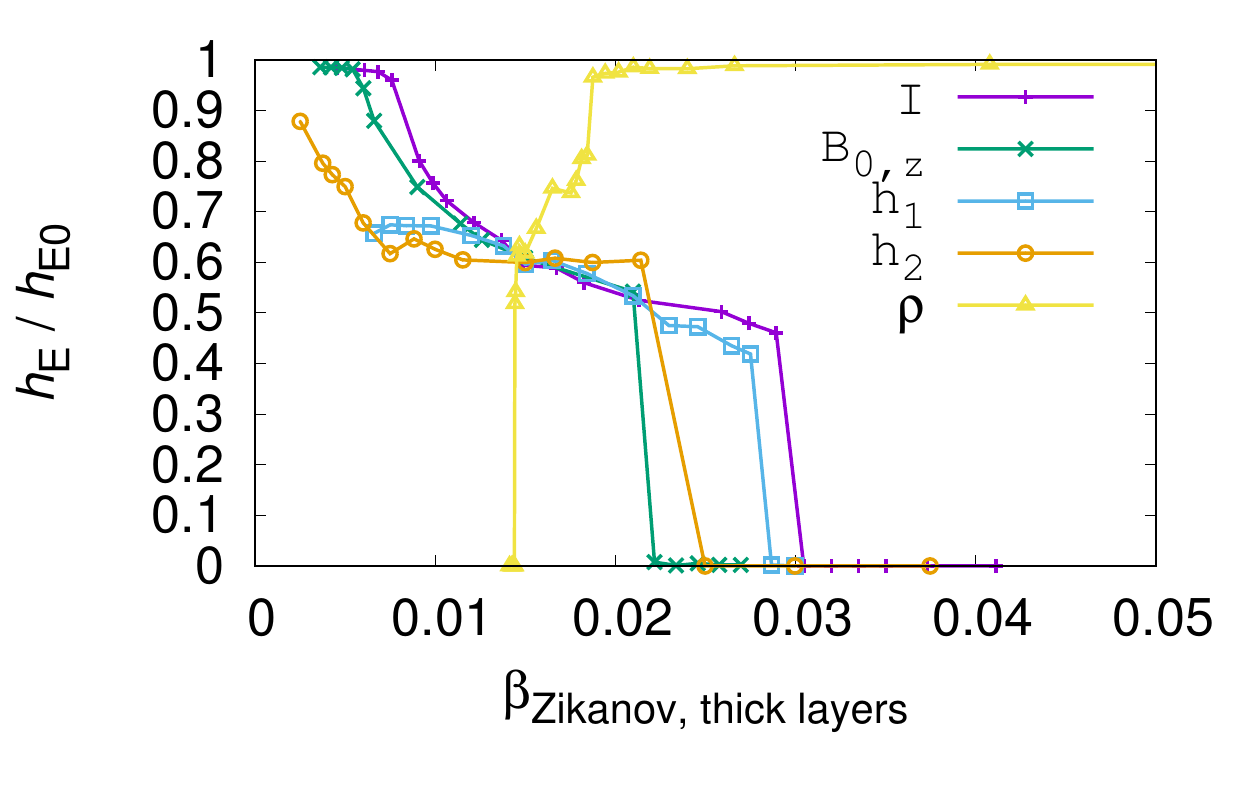}}\hfill
\caption{Remaining minimal salt layer thickness depending on
the dimensionless parameter $\beta$ as deduced from Davidson and
Lindsay \cite{Davidson1998} (a) and Zikanov \cite{Zikanov2015} (b-c).}
\label{f:beta:davidsonZikanov}
\end{figure}

In figures \ref{f:beta:sele} and \ref{f:beta:davidsonZikanov} we
illustrate the final height of the salt layer depending on $\beta$. The five
different curves represent a change of the cell current, magnetic
field, initial height of the upper metal or electrolyte and
density. In all diagrams we observe a good coincidence of
the curves for varying $I$ and $B_{0,z}$. A certain deviation
can be explained by the different damping nature of an increasing $I$
or $B_{0,z}$ \cite{Descloux1991}. 

The curves for changing the
height of the anode or electrolyte layer already deviate significantly. For
small $\beta$ especially the anode curve will \emph{not} converge to one, i.e. metal pad
rolling can appear in infinitely high cells. None of the dimensionless
parameters can correctly describe the onset of metal pad rolling
depending on the layer height; this is not surprising as almost all 
parameters were developed for shallow layers only. However, the region
of short circuit still seems to be described quite well, especially by
$\beta_\text{Sele}$. 

Finally, $\beta_\text{Davidson}$ and $\beta_\text{Zikanov}$
do not describe the influence of density acceptably, while
$\beta_\text{Sele}$ yields a good coincidence with the curves for $I$
and $B_{0,z}$. We will therefore focus on the original Sele
criterion in the following. At $\beta_\text{cr}\approx0.35$, metal pad
rolling appears for the first time; for lower values of $\beta$ the
cell is stable (for our aspect ratio). In contrast to linear stability
analysis which suggests that cylindrical cells are always unstable
\cite{Bojarevics1994} ($\beta_\text{cr} = 0$), our result indicate that
viscosity, induction and/or surface tension effects can shift the
threshold for onset of sloshing.

A second remarkable region in figure
\ref{f:beta:sele} is $\beta=2.0\dots3.2$ -- here the cell is
short-circuited very suddenly. This happens typically at a remaining
salt layer thickness of about $50$\% of the initial value. It is not
clear, whether $\beta>2$ or a certain salt layer thickness are
responsible for the sudden short-circuit. 

In summary, the Sele criterion $\beta$ allows quite well defining the
onset of metal pad rolling in LMBs -- but only for our fixed aspect
ratio. It is even possible to estimate the critical value for a
short-circuit of the cell (for any aspect ratio). An improved
dimensionless parameter should 
be developed for deep layers to better model the influence of the
layer thicknesses. It should further include viscosity, surface
tension as well as induction effects.

\subsection{Wave equation and short-circuit}\label{sec:short-cicuit}
\begin{figure}[t!]
\centering
 \subfigure[]{\includegraphics[width=0.45\textwidth]{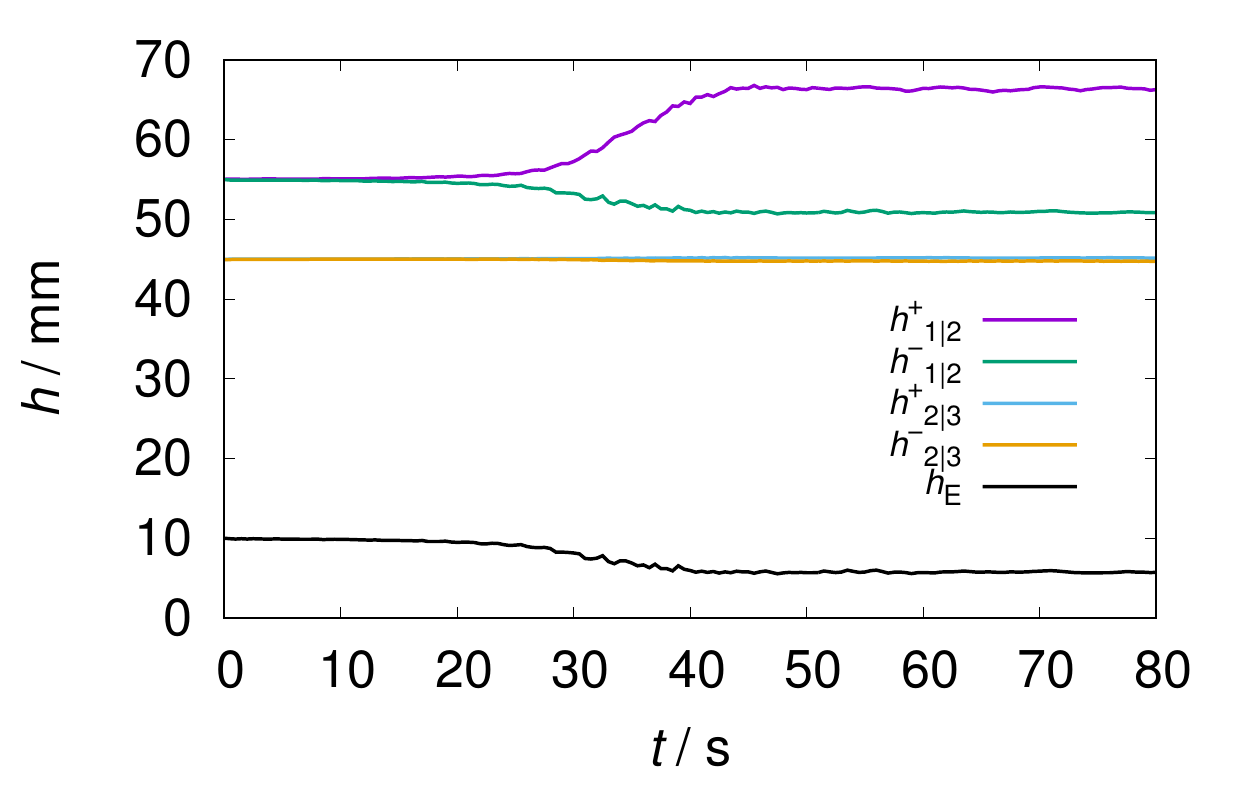}}
 \subfigure[]{\includegraphics[width=0.45\textwidth]{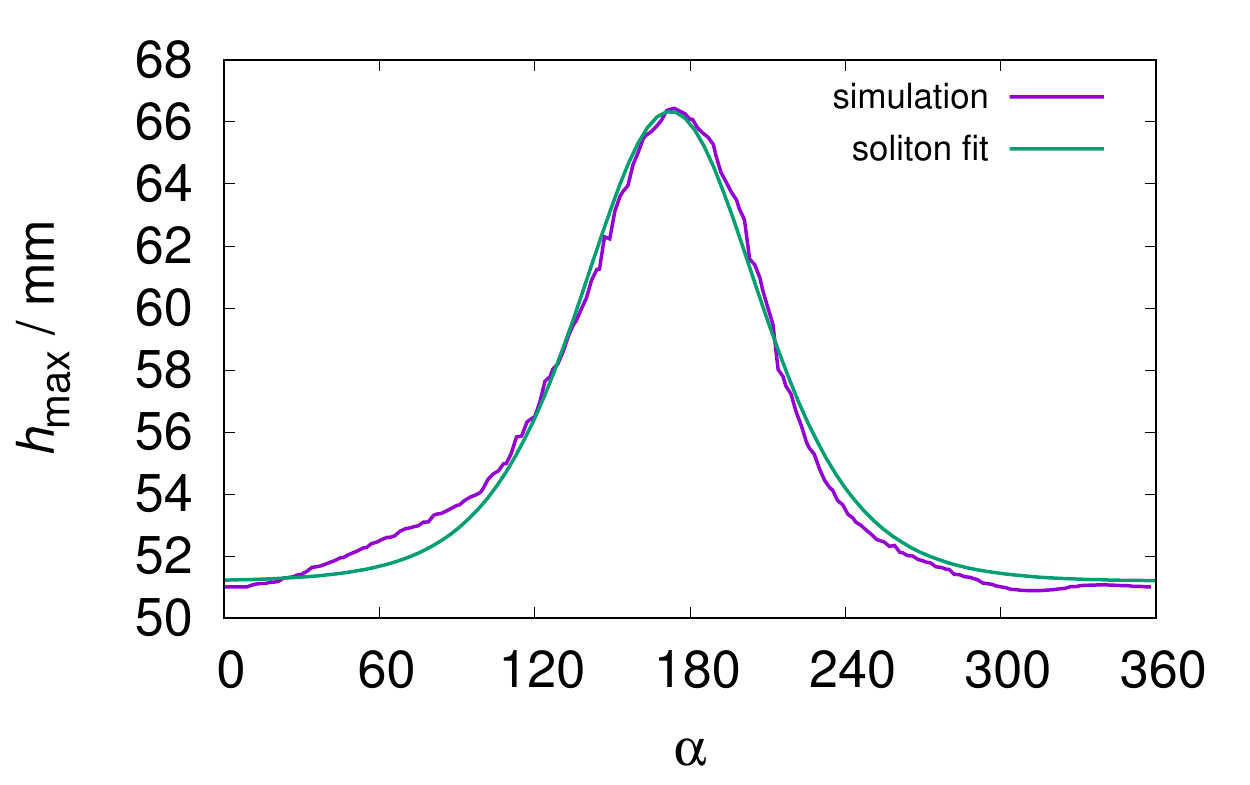}}
\caption{Temporal evolution of electrolyte layer thickness and minimal
and maximal elevation of both interfaces (a) and maximal elevation of
the upper interface over the circumference (b) ($I=100$\,A,
$B_z=10$\,mT, $h_1=h_3=4.5$\,cm, $h_2=1$\,cm).}
\label{f:soliton1}
\end{figure}

In this section we explore the shape of the interface as well as 
the sudden short-circuit. Figure \ref{f:soliton1}a shows the minimal and
maximal height of both interfaces -- depending on time. We observe
after $50$\,s a stationary sloshing; the interfaces do not touch each
other. Figure \ref{f:soliton1}b now illustrates the shape of the
upper metal-electrolyte interface around the circumference of the
cylinder. We assess here the hypothesis that this shape can be
described as a solitary wave \cite{Munteanu2004}. Similar to the
solution of the Korteweg-de-Vries equation, we describe the interface
height as
\begin{equation}
h_I = h_0 + a \sech^2\left(kr(\alpha -\omega t)\right)
\end{equation}
with its minimal height $h_0$, amplitude $a$ and angle $\alpha$. The
wave number $k$ defines the width of the crest, the angular frequency
$\omega$ its speed. This equation nicely fits the observed interface
shape (figure \ref{f:soliton1}b); however, it allows only
symmetrical crests. Especially for high currents the real wave front
is very steep, but the tail quite smooth.

\begin{figure}[b!]
\centering
 \subfigure[]{\includegraphics[width=0.45\textwidth]{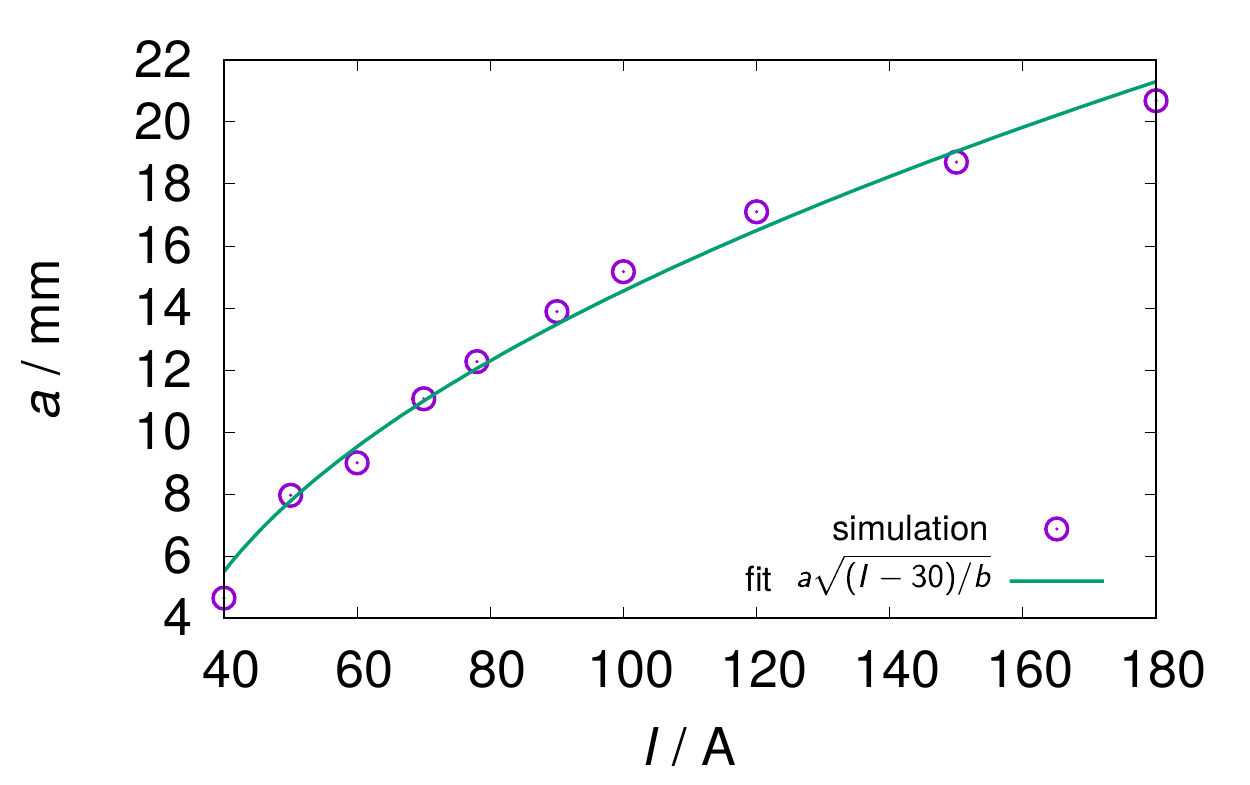}}
 \subfigure[]{\includegraphics[width=0.45\textwidth]{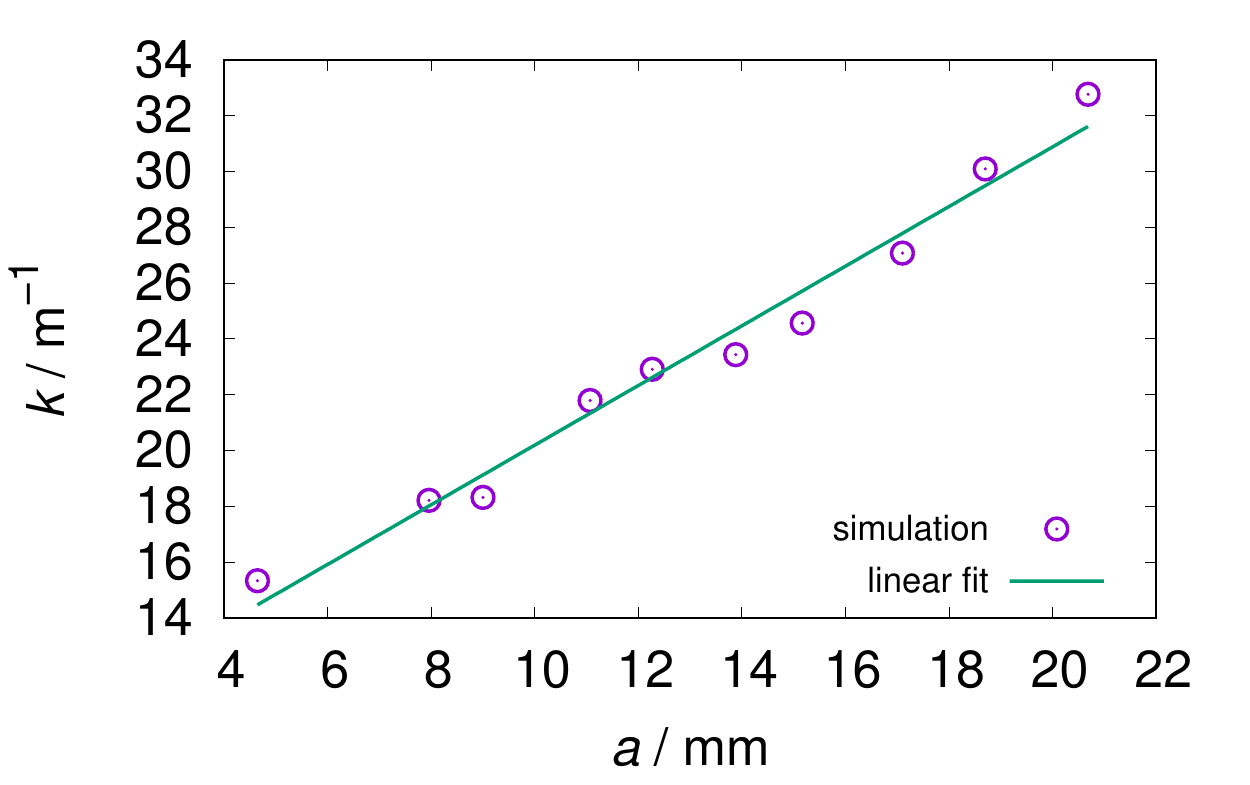}}
\caption{Amplitude of the elevation of the upper interface depending
  on the cell current (a) and dependence of the wave number on the
  amplitude of the wave (b) ($B_z=10$\,mT, $h_1=h_3=4.5$\,cm,
  $h_2=1$\,cm).}
\label{f:solitonEqn}
\end{figure}
In figure \ref{f:solitonEqn}a we show the dependence of the wave
amplitude $a$ on the cell current $I$. As the relation between
both quantities appears to be a square root curve, Ginzburg-Landau
theory may apply here\cite{Aranson2002}. However, the latter is 
usually used to decribe weakly nonlinear regimes, while solitons
are strongly nonlinear solutions.

Further, we study the relation of width and
amplitude of the wave crest by comparing $k$ and $a$ in figure
\ref{f:solitonEqn}b. The relation is linear, which does not match well
the solution of the Korteweg-de-Vries equation. Its solution usually
suggests a quadratic amplitude compared to the wave number
\cite{Munteanu2004}. 

\begin{figure}[t!]
\centering
 \includegraphics[width=0.55\textwidth]{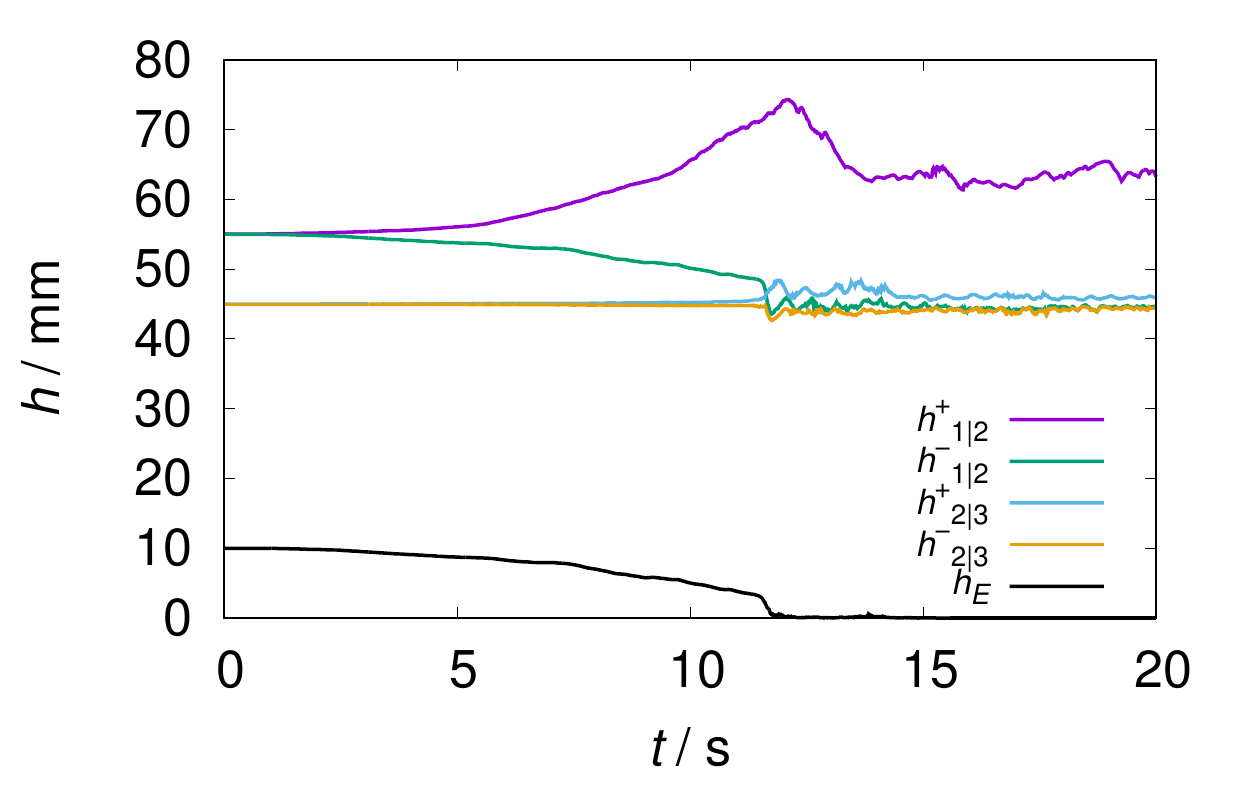}
\caption{Temporal evolution of electrolyte layer thickness and minimal
  and maximal elevation of both interfaces for $I=200$\,A
  ($B_z=10$\,mT, $h_1=h_3=4.5$\,cm, $h_2=1$\,cm).}
\label{f:minmax}
\end{figure}

\begin{figure}[t!]
\centering
 \subfigure[\,$t=10.7$\,s]{\includegraphics[width=\textwidth]{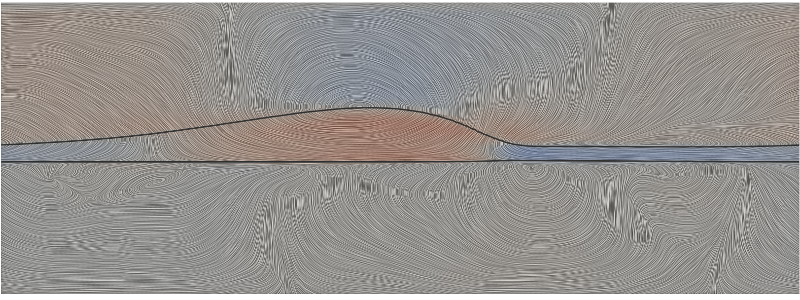}}
 \subfigure[\,$t=12.47$\,s]{\includegraphics[width=\textwidth]{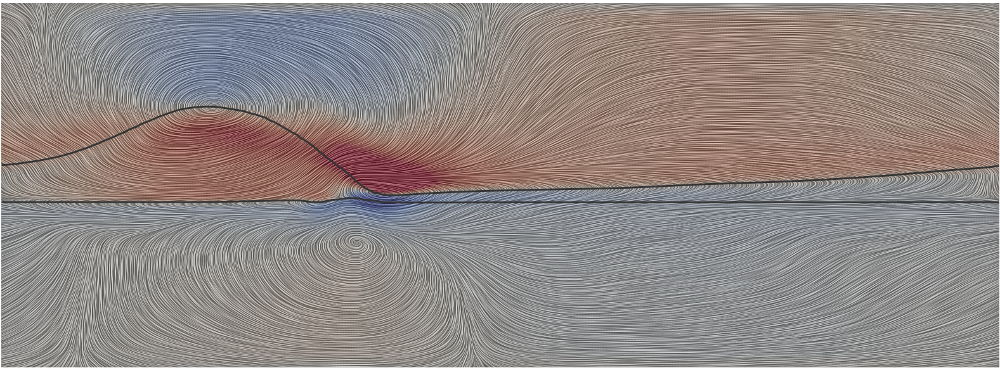}}
 \subfigure[\,$t=12.67$\,s]{\includegraphics[width=\textwidth]{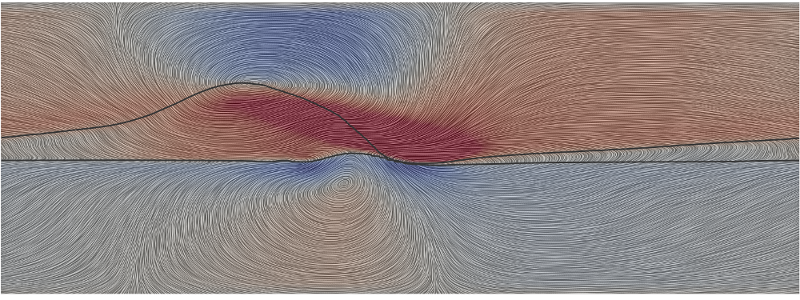}}
\caption{Interface deformation, velocity and streamlines on 
  the curved surface area of the cylinder ($I=200$\,A, $B_z=10$\,mT,
  $h_1=h_3=4.5$\,cm, $h_2=1$\,cm). Red color indicates a positive
  velocity (to the right, i.e. anti-clockwise), blue color a negativ
  flow (to the left, i.e. clockwise).}
\label{f:umfang}
\end{figure}

In figure \ref{f:minmax} we show a simulation of a sloshing
instability leading to a short-circuit. The trough decreases quite
slowly, and short-circuits then suddenly. At the same time, the lower
interface is deformed and starts to oscillate; the crest of the upper
interface decreases again.
This sudden interface deformation can have three reasons: surface
tension, electromagnetic force or velocity. Surface tension rather
dampens waves; the Lorentz force at the short-circuit will point to
the cell axis and decrease pressure at the wall. Both do not provide a clear
explanation of the short-circuit. We show therefore in figure
\ref{f:umfang} three different plots of the velocity on the
circumference of the cell. With a rising crest, the slope of the wave
front increases. This leads to a considerable flow in front of it in
anti-clockwise direction (figure \ref{f:umfang}b). We assume that this flow
decreases the local pressure leading to a sudden pinching of the
electrolyte layer (figure \ref{f:umfang}c), maybe in concert with the
locally concentrated Lorentz force. The lowered pressure may also
explain the waves appearing at the lower interface.

\clearpage
\subsection{Conclusion and application to real cells}
Today's liquid metal batteries (LMB) are rather shallow, having a diameter in
the order of $20$\,cm. Next generation LMBs may hopefully be
considerably larger in size, with the height depending on the desired
capacity of the cell. For small as for large cells, the
electrolyte layer must be as thin as possible due to its high
resistance. A typical value\cite{Weber2016} is 4-10\,mm. Current
densities can strongly vary\cite{Weber2016}: from $100$\,mA/cm$^2$ for energy-efficient
discharge and high rate capabilities
to $1$\,A/cm$^2$ for fast (dis-)charge. Li$||$Te and Li$||$Se cells
even reached values of $13$\,A/cm$^2$. A realistic
value of the background magnetic field can be as for aluminium
reduction cells between 1 and $10$\,mT\cite{Molokov2011}.
\begin{figure}[b!]
\centering
\includegraphics[width=0.9\textwidth]{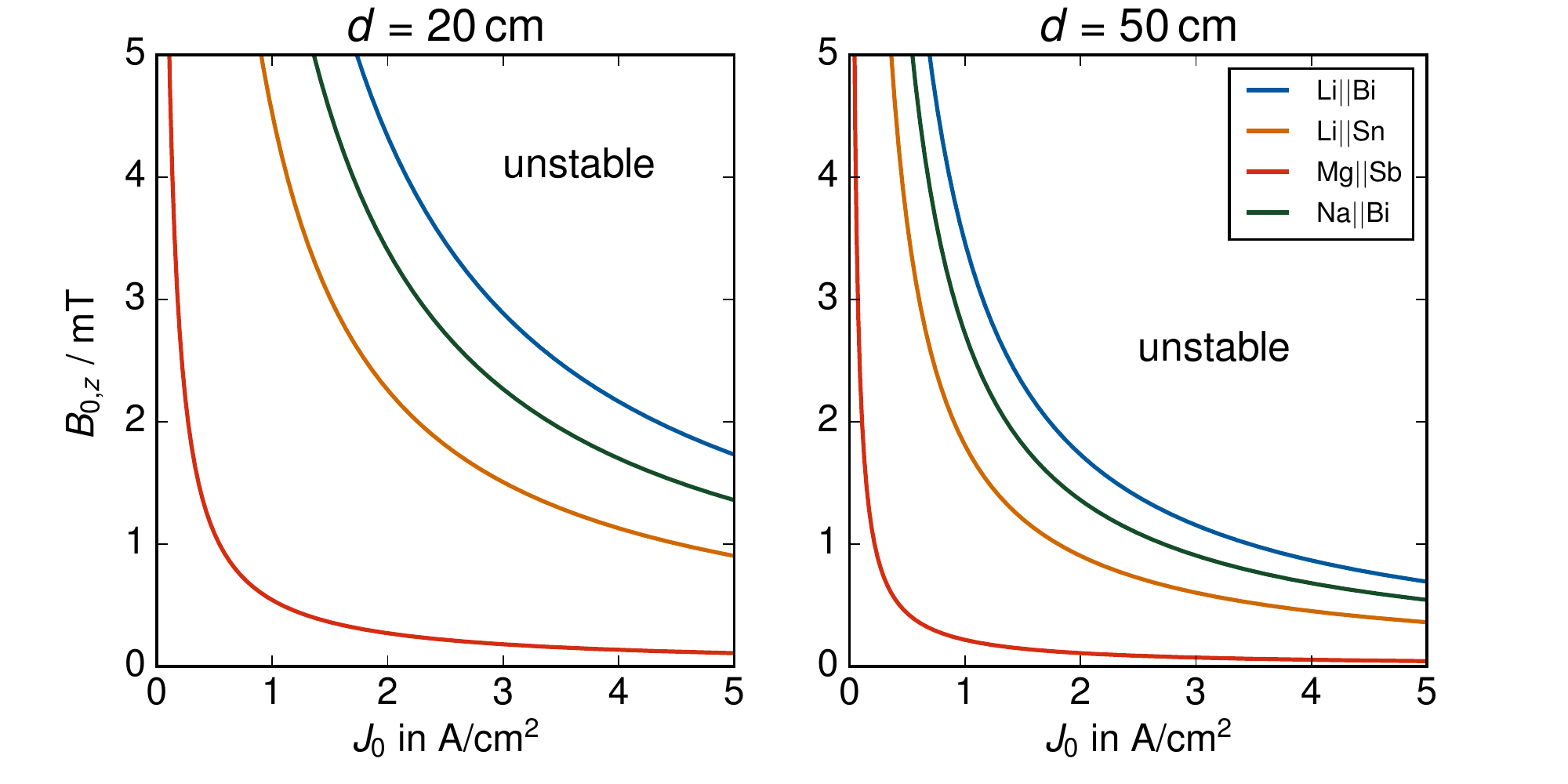}
\caption
{Onset of sloshing ($\beta=0.35$) depending on current density $J_0$
  and magnetic background field $B_{0,z}$. The aspect ratio of the
  anode $h_1/d = 0.45$ is constant, the electrolyte layer 4\,mm
  thick.}
\label{f:onset}
\end{figure}

Using the Sele criterion
\begin{equation}\label{eqn:sele:final}
\beta_\text{Sele} = \frac{Jd^2\pi b}{4g(\rho_2-\rho_1) h_1h_1} <
\beta_\text{cr}
\end{equation}
and knowing the critical (Sele) parameter for the onset of sloshing
$\beta=0.35$ and the short-circuit $\beta=2$,
we can estimate whether a certain LMB will be stable, unstable or even
short-circuited. The necessary physical quantities of the most common
LMBs are listed in table \ref{t:lmbs}. Note that strictly speaking
$\beta=0.35$ for the onset of sloshing holds only for our aspect
ratios ($h_1/d = 0.45$, $h_2/d = 0.1$). Only to get a first impression
we \emph{assume} $\beta_\text{cr}$ to be the same also for shallower 
electrolyte layers. We show in figure \ref{f:onset} the onset of
sloshing depending on current density and $B_{0,z}$ for different
cells of diameter 20 and 50\,cm. Obviously, metal pad rolling can
already set in in rather small cells with a diameter of a few
decimeters. Due to its small density difference, the Mg$||$Sb cell is
the most vulnerable one.
\begin{figure}[t!]
\centering
 \includegraphics[width=\textwidth]{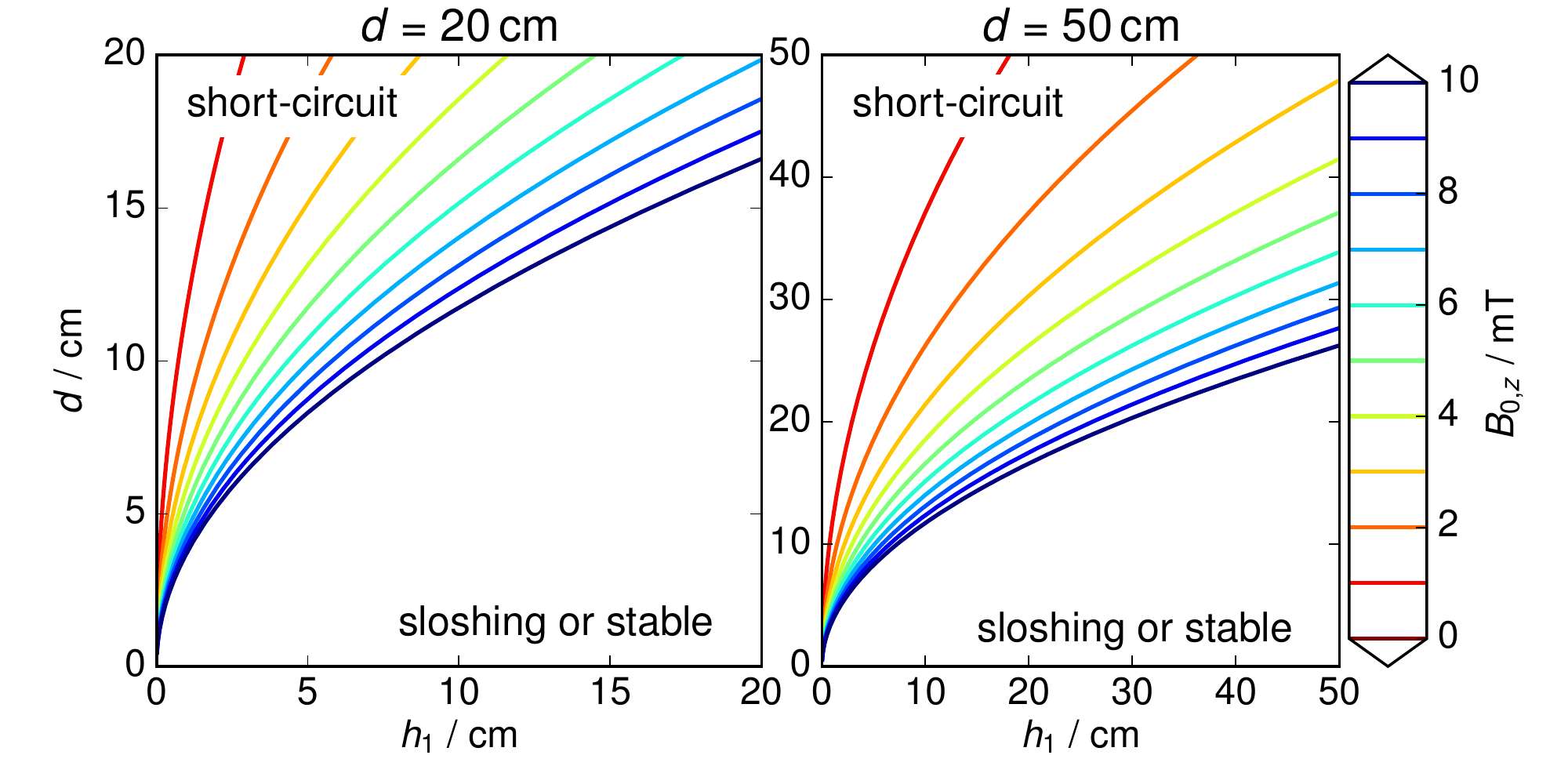}
\caption
{ Short circuit ($\beta=2$) of an LMB
  depending on the height of the top metal layer $h_1$, the cell
  diameter $d$ and magnetic field $B_{0,z}$ for a Mg$||$Sb cell. The
  height of the electrolyte layer is assumed to be 4\,mm, the current
  density 1\,A/cm$^2$.}
\label{f:short}
\end{figure}

In figure \ref{f:short} we illustrate the short-circuit of our
exemplary Mg$||$Sb cell in dependence of $B_{0,z}$, the diameter $d$
and upper metal height $h_1$. We use an electrolyte thickness of 4\,mm
and a current density of $1$\,A/cm$^2$. A small 10\,cm Mg$||$Sb cell,
using a 5\,cm high upper metal layer can already be short-circuited by
the presence of a 6\,mT strong vertical background field.
\begin{table}[h!]\vspace{8pt}
\centering
\caption{Properties of liquid metal batteries: operating temperature,
  open circuit voltage, maximum current density and density difference
  upper metal-electrolyte.}\label{t:lmbs}
\begin{tabular}{llrrrrrrrrrrrrrrrr}\hline
\multicolumn{1}{c}{cell}&\multicolumn{1}{c}{electrolyte} & \multicolumn{1}{c}{$T_{op}$} & \multicolumn{1}{c}{$U_0$} & \multicolumn{1}{c}{$J_{max}$} &
\multicolumn{1}{c}{$\Delta\rho$} & \multicolumn{1}{c}{literature}\\
%\cmidrule(lr){3-3}\cmidrule(lr){4-4}\cmidrule(lr){5-5}\cmidrule(lr){6-6}
\hline
&&\multicolumn{1}{c}{$^\circ$C} & \multicolumn{1}{c}{V} & \multicolumn{1}{c}{A/cm$^2$} & \multicolumn{1}{c}{kg/m$^3$}\\
\hline
%Ca$||$Bi & CaCl$_2$-LiCl-NaCl & 550 & 1.2 & 2 &1434&&9720&4.0&&0.7&24&&1.2&0.39&&0.34&\cite{Kim2013a,Ouchi2016,Sobolev2010}\\ Unfug, schmiltz erst bei 842°C
K$||$Hg & KBr-KI-KOH          & 250 & 1.1  & 0.1 &1759&\cite{Agruss1962,Chum1980,IAEA2008,Gale2004,Karas1963}\\
Li$||$Bi & LiCl-LiF-LiI       & 485 & 1.4  & 6.1 &2202&\cite{Shimotake1969,Sobolev2010,Zinkle1998}\\
Li$||$Pb & LiCl-LiF-LiI       & 483 & 0.6  & 0.4 &2202&\cite{Chum1981,Wang2014,Zinkle1998,Sobolev2007}\\
%Li$||$Sb & LiCl-LiF-LiI & \\
Li$||$Se & LiCl-LiF-LiI       & 375 & 2.4  & 13  &2192&\cite{Cairns1969b,Zinkle1998,Gale2004}\\
Li$||$Sn & LiCl-LiF           & 400 & 0.75 & 1   &1149&\cite{Cairns1967,Swinkels1971,Zinkle1998,Gale2004,Lyon1954}\\
Li$||$Te & LiCl-LiF-LiI       & 475 & 1.9  & 12.7&2201&\cite{Cairns1967,Cairns1969c,Zinkle1998,Gale2004}\\
Li$||$Zn & KCl-LiCl           & 486 & 0.64 & 0.3 &1140&\cite{Chum1981,Zinkle1998,Gale2004}\\
Mg$||$Sb & KCl-MgCl$_2$-NaCl  & 700 & 0.6  & 0.2 &138&\cite{Bradwell2011,Bradwell2012,Gale2004,Aqra2011}\\
Na$||$Bi & NaCl-NaI-NaF       & 550 & 0.7  & 2.2 &1729&\cite{Cairns1967,Sobolev2010,IAEA2008}\\
Na$||$Hg & NaI-NaOH           & 275 & 0.78 & 0.36&&\cite{Heredy1967,Spatocco2014,IAEA2008,Gale2004}\\
Na$||$Pb & NaCl-NaF-NaI       & 575 & 0.5  & 0.2 &1713&\cite{Chum1981,IAEA2008,Sobolev2010,Sobolev2007}\\
Na$||$Sn & NaCl-NaI           & 625 & 0.55 & 0.77&1554&\cite{Weaver1962,IAEA2008,Sobolev2010,Gale2004,Janz1979a}\\
\hline\end{tabular}\vspace{3pt}\end{table}

\section{Summary and outlook}
The main purpose of this paper was to show that the presence of a
vertical magnetic field can spark the metal pad roll 
instability in liquid metal batteries (LMBs). This interface instability
can appear in any cell as long as the current and magnetic background 
field are strong enough; it may finally short-circuit an LMB. In real cells (with a
limited current density) the appearance of sloshing must be taken into
account if the diameter is larger than some centimeters, especially for
Mg$||$Sb cells. Metal pad rolling can therefore be considered as one of the most 
important instabilities in the operation of LMBs. Yet, it can be prevented
by choosing high (upper metal) layers, by an appropriate design of the
bus system (minimising vertical magnetic fields) and by using a rectangular cross
section instead of  cylindrical or square cells.

Metal pad rolling is already well known from aluminium reduction cells. We
have indentified a Sele mechanism explaining the wave propagation. The
wave period is well described by the dispersion relation for gravity waves,
if accounting also for surface tension.
We have further studied a wide range of parameters influencing onset and
intensity of sloshing: besides of strong vertical fields, also high
cell currents lead to instability. Consequently, large cells are
particularly vulnerable. The density difference between alkaline
metal and salt should be high for stable LMBs; shallow (upper metal and
electrolyte) layers promote instability. High viscosities again 
stabilise the cell. Many of these parameters can be
combined to a dimensionless number (Sele criterion, equation
\ref{eqn:sele:final}) characterising onset and intensity of
sloshing. While the Sele criterion models well the influence of 
current, magnetic field and density, it partially fails describing the influence of
the layer thickness of typical LMBs. 

Increasing e.g. the cell current, the electrolytes minimal
thickness decreases (figure \ref{f:flow}). We have proposed to describe
 the influence of several parameters on this salt layer 
 thickness by Ginzburg-Landau theory. For small and moderate
 deformation of the electrolyte layer, we suggest to describe the
 shape of the upper metal-electrolyte interface as some kind of solitary
 wave. The sudden short-circuit of the salt layer is attributed
 to a strong flow in front of the wave crest.

Our simulation can only give a first rough overview on the sloshing
instability in LMBs. Maybe the most important step would be an
experimental evidence of the instability in a three layer
system. Although metal pad rolling was intensively studied in (the
two-layer layer system of) aluminium reduction cells, the dimenionless
parameter defining its onset is imperfect. Most importantly, the Sele
criterion should be extended to cells of high aspect ratio to describe
the influence of the layer heights better.
It could further be complemented by the influence of induction,
surface tension and viscosity and maybe adapted to our three layer
system. The very sudden short-circuit will need further and
more detailed studies. For this purpose it may be necessary to improve
the mixture model for conductivity as well as the implementation of
surface tension in our solver. A comparison of the OpenFOAM solver with the
spectral code SFEMaNS \cite{Guermond2009} is planned.

\section*{Acknowledgments}
This work was supported by Helmholtz-Gemeinschaft Deutscher
Forschungs\-zentren (HGF) in frame of the Helmholtz Alliance 
``Liquid metal technologies'' (LIMTECH). The
computations were performed on the Bull HPC-Cluster 
``Taurus'' at the Center for Information Services and 
High Performance Computing (ZIH) at TU Dresden and on the 
cluster ``Hydra'' at Helmholtz-Zentrum Dresden -- Rossendorf.
 Fruitful
discussions with Valdis Bojarevics,
Douglas Kelley, Cornel Lalau, Steffen Landgraf, Michael Nimtz, Marco Starace and Oleg Zikanov
on several aspects of metal pad roll instability and liquid metal 
batteries are gratefully acknowledged. N. Weber thanks Henrik
Schulz for the HPC support.

\bibliography{literatur}

\appendix

\section{Numerical model}
\begin{figure}[htb!]
\centerline{
\includegraphics[width=\columnwidth]{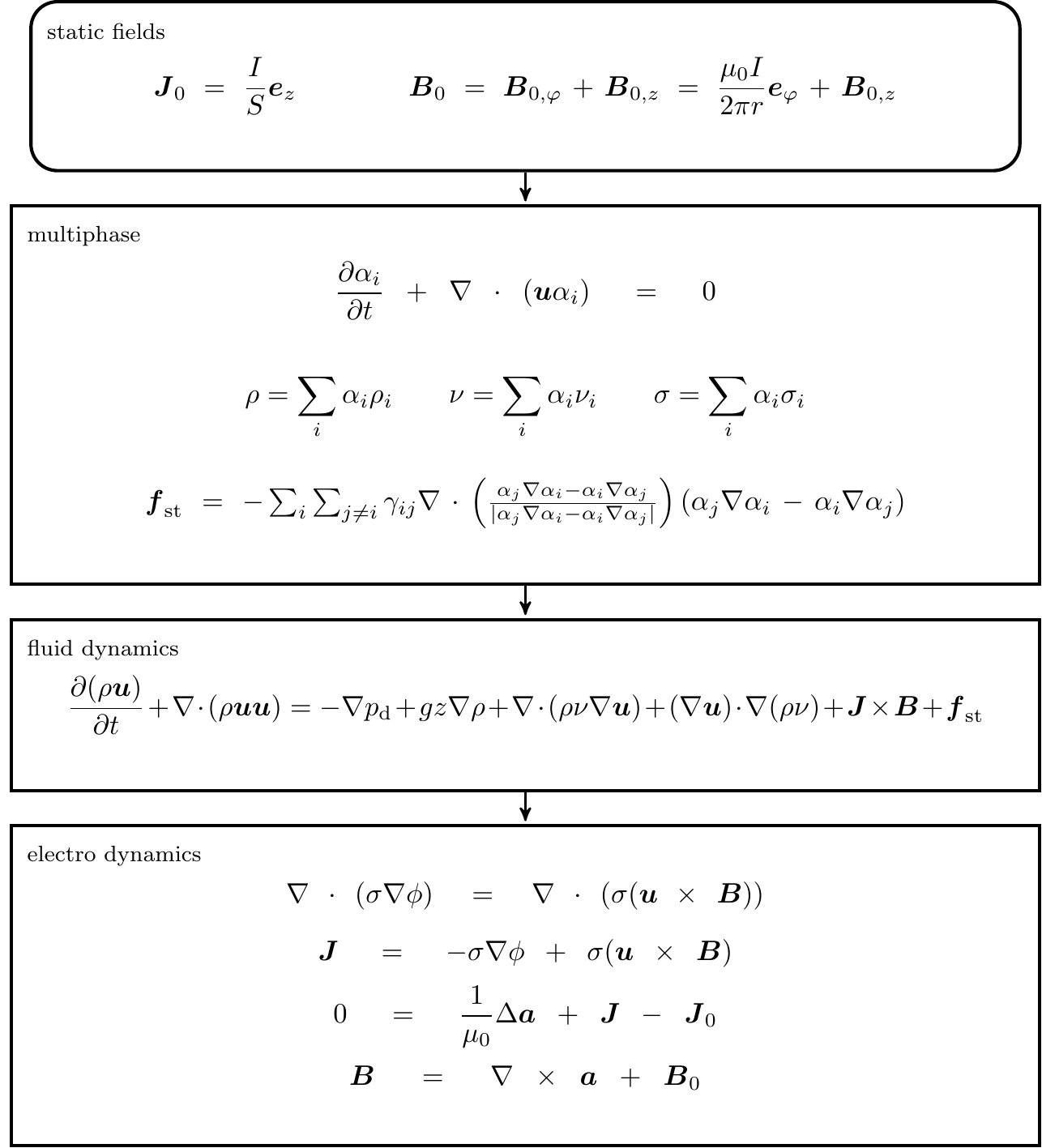}}
\caption{Flowchart for multiphase simulation.}
\label{f:workflow}
\end{figure}

\end{document}